\documentclass[twocolumn,showpacs,amsmath,prb,superscriptaddress]{revtex4-2}
\usepackage{epsfig,psfrag}
\usepackage{amsbsy}
\usepackage{amsfonts}
\usepackage{bbm}
\usepackage{tabularx}
\usepackage{euscript}
\usepackage{amsfonts}
\usepackage{exscale}
 \usepackage{slashed}
\usepackage{comment}
\usepackage{amsmath,amssymb}
\usepackage{graphicx}
\usepackage{dcolumn}
\usepackage{bm}
\usepackage{braket}
\usepackage{verbatim}
\usepackage{float}
\usepackage{bbold}
\usepackage{color}
\usepackage{xcolor}
\usepackage{relsize}
\usepackage{amsthm}
\usepackage{enumerate}
\usepackage{soul,xcolor}
\usepackage[normalem]{ulem}
\usepackage[T1]{fontenc}
\usepackage[colorlinks=true ,urlcolor=blue,urlbordercolor={0 1 1}]{hyperref}

\begin{document}
\graphicspath{{figures/}}

\setstcolor{red}

\title{Magnetic Breakdown and Anomalous Quantum Oscillation in Rhombohedral Tetralayer Graphene }

\author{Jing-Yu Zhao}
\affiliation{
Condensed Matter Theory Center and Joint Quantum Institute,
Department of Physics, University of Maryland, College Park, Maryland 20742, USA}
\author{Yang-Zhi Chou}
\affiliation{
Condensed Matter Theory Center and Joint Quantum Institute,
Department of Physics, University of Maryland, College Park, Maryland 20742, USA}
\author{Sankar Das Sarma}
\affiliation{
Condensed Matter Theory Center and Joint Quantum Institute,
Department of Physics, University of Maryland, College Park, Maryland 20742, USA}

\date{\today}

\begin{abstract}
We investigate magnetic breakdown near Van Hove singularities (VHSs) in the electron-doped rhombohedral tetralayer graphene,   
where chiral superconductivity has recently been reported. Using the noninteracting band structure and Kubo formula, we identify anomalous Shubnikov–de Haas effects: Ring-like structures in the Landau fan and anomalous high-frequency peaks in the frequency spectra. 
These anomalous quantum oscillations can be understood by the reconstruction from magnetic breakdown among three nearby Fermi pockets separated by VHSs.
Remarkably, these qualitative anomalous features persist into a stronger-VHS regime, where the semiclassical picture breaks down. 
The temperature and (weak) disorder dependence of the oscillations are also investigated. Our results establish that the magnetic-breakdown-induced anomalous quantum oscillation provides a general distinctive probe for the underlying Fermi-surface geometry associated with VHSs and may explain the recent quantum oscillation experiment in rhombohedral tetralayer graphene [\href{https://arxiv.org/abs/2606.05356}{arXiv:2606.05356}]. 

\end{abstract}
\maketitle{}

\begin{figure}[t]
    \centering
    \includegraphics[width=0.95\linewidth]{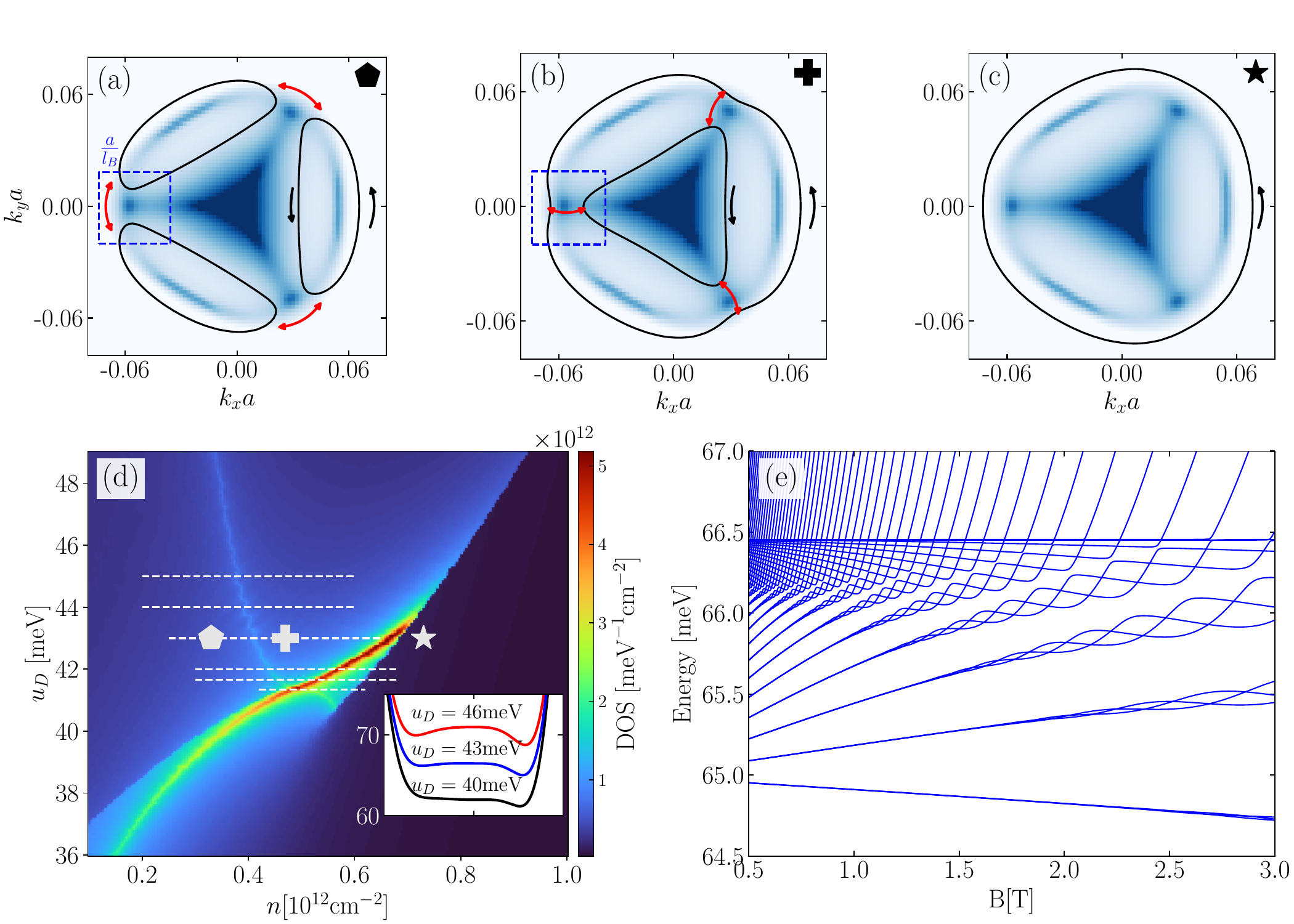}
    \caption{
    (a)--(c) Representative Fermi surfaces and possible magnetic breakdowns at a displacement field of $u_D=43\mathrm{meV}$ and electron densities of 
    (a) $n=0.33\times10^{12}~\mathrm{cm}^{-2}$, (b) $n=0.47\times10^{12}~\mathrm{cm}^{-2}$, and (c) $n=0.73\times10^{12}~\mathrm{cm}^{-2}$.
    The blue color scale represents the momentum-resolved density of states on a logarithmic scale.
    In a magnetic field, quasiparticles undergo semiclassical cyclotron motion along the Fermi surfaces, as indicated by the black arrows. 
    They may also tunnel between neighboring Fermi-surface pockets through magnetic breakdown, as indicated by the red arrows in (a) and (b).
    The inverse magnetic length $a/l_B$ at $B=1~\mathrm{T}$ is indicated by the blue squares in (a) and (b).
    (d) Density of states of R4G as a function of carrier density $n$ and displacement field $u_D$.
    The parameter points corresponding to panels (a)--(c) are marked in the diagram.
    The white dashed lines mark the density regimes where at least one of the ``multitone'' frequencies is observable. 
    Typical band structures at $u_D=40\,\mathrm{meV}$, $u_D= 43\,\mathrm{meV}$ and $u_D=46\,\mathrm{meV}$ are shown in the inset. 
    (e) Landau-level energies as a function of $B$ for $u_D = 43\,\mathrm{meV}$.  
    }
    \label{fig:FS}
\end{figure}

\textit{Introduction---}
The recent observation of putative quarter-metal superconductivity in rhombohedral tetralayer and pentalayer graphene \cite{HanRnGSC2025b} has established rhombohedral $n$-layer graphene (R$n$G) as a clean and tunable platform for unconventional superconductivity \cite{HanRnGSC2025b, JIAli2025,kalantreFermiology2026, huaR6G2026}. 
Remarkably, the superconductivity with hysteretic magnetotransport has been interpreted as evidence for broken time-reversal symmetry and a spin-valley polarized normal state, indicating a possible chiral superconductivity (CSC).
The pairing mechanisms and topological properties of the CSC from a single-flavor metal have become active topics of research \cite{geierCSC2025,chouR4GSC2025,yangPDW2025,parraRPA2025, gaggioliVAV2025, 
chouQM2026, christosPDW2025, sauAnomalousHall2024, zhuCSC2026, gilPDW2026, dongQM2026, qinR4G2026,  
chenDiode2025,jahinKohnLuttinger2026, liBerry2025,maymannBerry2026, patriRnG2025, wangCSC2026, yoonQMl2026a}. 
The resulting superconducting instability depends sensitively on the geometry and topology of the underlying normal-state Fermi surface.  
However, the Fermi surface is known to host intricate Van Hove Singularities (VHSs) and Fermi surface geometry, which may become further complicated in the presence of interactions. 
It is therefore essential to understand the normal-state electronic structure and identify its experimentally accessible signatures.

Recent quantum-oscillation measurements \cite{kalantreFermiology2026} provide an important probe of the normal state underlying and surrounding CSC in R4G.
The measured Shubnikov-de Haas (SdH) oscillations exhibit a complex ``multitone'' structure that cannot be explained by a straightforward application of the Onsager relation to a spin- valley-polarized metal. 
In particular, two high-frequency components persist throughout the superconducting regime, each corresponding to an effective carrier density much larger than the nominal gate-defined density $n$. 
These observations rule out a simple, singly connected Fermi surface, while the nature of the underlying normal state remains unresolved. 
Thus, it is essential to investigate the SdH oscillation in a nearly flat band with complex Fermi surfaces, incorporating full quantum mechanical effects (i.e., beyond semiclassical treatment). 

In this letter, we systematically study the quantum tunneling-induced magnetic breakdown
\cite{pippard1962a,pippard1964,chambersMB1966, alexandradinataMB2017, alexandradinataModernMB2018, starkMB1971a} 
near the VHSs in electron-doped R4G and show its consequence on the SdH oscillations. 
The relevant Fermi surface geometry is illustrated in Fig.~\ref{fig:FS}(a), where the original degenerate Landau levels of three small electron pockets are intertwined by the magnetic field-induced tunneling across the VHS. 
The magnetic breakdown generates additional crossings and ring-like structures in the Landau fan, ultimately producing the two non-Onsager high-frequency components in the quantum oscillations. 
Although magnetic breakdown has been studied for decades, including in moir\'e materials \cite{luMBTBG2014,hejaziTwisted2019, paulMoireLandauFans2022,paulMB2024, bocarslyHaasVanAlphen2024}, its consequences are usually understood in terms of combination frequencies derived from the underlying Onsager frequencies. 
The mechanism proposed here goes beyond this conventional picture and may provide a distinctive transport probe of similar Lifshitz transitions from an annular Fermi surface to multiple disconnected pockets.

We calculate the longitudinal resistivity and the corresponding SdH Fourier spectra using the Landau-level spectrum of the noninteracting R4G Hamiltonian \cite{kalantreFermiology2026, auerbachIsospin2025} and the standard Kubo formula for transport \cite{ando1974,ando1974a,andoRMP1982a}. 
We obtain two nearly density-independent SdH frequencies with $n_{\mathrm{SdH}}>n$ that are robust to thermal broadening, consistent with the experimentally observed ``multitone'' frequencies\cite{kalantreFermiology2026}. 
These qualitative anomalous features persist into a stronger-VHS regime, where the semiclassical picture breaks down. We also comment on the seemingly coincidental relation between ``multitone'' features and CSC. 
Our results indicate that the noninteracting band description, without invoking additional symmetry breaking or strong correlation, 
already captures the essential fermiology of the spin- and valley-polarized metallic state in the recent R4G experiment \cite{kalantreFermiology2026}.

\textit{Fermi surface of R4G and Magnetic breakdown ---}
We study the R4G (i.e., an ABCA stacking) through the nonintearacting $k\cdot p$ continuum Hamiltonian with parameters taken from \cite{kalantreFermiology2026, auerbachIsospin2025}, 
\begin{equation} \label{eqn:HR4G}
    H_0 = \sum_{\eta=\pm1}\sum_{\mathbf{k}}
    \Psi_{\eta,\mathbf{k}}^{\dagger}
    h_{\eta}(\mathbf{k})
    \Psi_{\eta,\mathbf{k}},
\end{equation}
where $\Psi_{\eta,\mathbf{k}}$ is an eight component spinor in the basis  $(A_1,B_1,A_2,B_2,A_3,B_3,A_4,B_4)$, $\eta = \pm 1$ labels the valley $K$ and $K'$. 
Throughout this work, we assume a fully spin- and valley- polarized state in the $K$ valley. 
Details of the matrix $h_\eta(\mathbf{k})$ and the corresponding parameters are provided in the Supplemental Material (SM) \cite{SM}. 
Rhombohedral ABCA stacking produces a low-energy conduction band with a large effective mass \cite{minABC2008,koshinoTrigonal2009}, which is further flattened by a perpendicular displacement field $u_D$, defined as the potential difference between adjacent layers. 
In the regime relevant to CSC, the band develops a Mexican-hat dispersion and multiple VHSs.   

First, we discuss the Fermi surface evolution at $u_D=43~\mathrm{meV}$, with representative Fermi surfaces shown in Fig.~\ref{fig:FS}. 
At high electron density as in Fig.~\ref{fig:FS}(c), the occupied states form a single, simply connected electron pocket whose area equals the carrier density. 
Upon lowering the density, the system undergoes two successive Lifshitz transitions, corresponding to two VHSs.
First, a hole-like contour emerges near the center of the electron pocket, giving rise to an annular Fermi sea as in Fig.~\ref{fig:FS}(b). 
As density further decreases, trigonal warping splits the annular Fermi surface into three disconnected electron pockets related by $C_3$ symmetry as shown in Fig.~\ref{fig:FS}(a). 
The density of states as a function of carrier density $n$ and displacement field $u_D$ is shown in Fig.~\ref{fig:FS} (d), where the two sets of VHSs can be clearly seen. 
Lowering the value of $u_D$, the region of annular Fermi surfaces shrinks, and two VHSs merge to one at around $u_D=41.7$ meV, producing the largest density of states.

The VHSs in R4G are believed to be crucial for the formation of CSC \cite{chouR4GSC2025,gilPDW2026}.
Here we show that these VHSs can also cause drastic changes in the SdH frequencies for a large magnetic field $B$. 
In a weak perpendicular magnetic field, a semi-classical wave packet approximately follows a constant-energy contour indicated by the black arrows in Fig.~\ref{fig:FS} (a)--(c). 
Each isolated pocket is quantized independently according to the Onsager--Bohr--Sommerfeld condition \cite{shoenbergSdH1984a}
\begin{equation}\label{eqn:OBS}
    l_B^2 A_p(E_N)=2\pi\bigl(N+\gamma_p(E_N)\bigr),
    \qquad
    l_B=\sqrt{\frac{\hbar}{eB}},
\end{equation}
where $N=0,1,\ldots$ denotes the Landau level index, $A_p(E)$ is the momentum-space area enclosed by one pocket $p$ at energy $E$, and $\gamma_p$ includes the Maslov, orbital magnetization and geometric-phase corrections \cite{shoenbergSdH1984a, mikitikBerrysPhase1999}.

The independent-orbit description breaks down near the VHSs. 
Since the kinetic momenta $\hat{\pi}_i=-i\partial_i+eA_i$ satisfy
\begin{equation}
    [\hat{\pi}_x,\hat{\pi}_y]=-{i}/{l_B^2}~,
\end{equation}
Eq.~\eqref{eqn:OBS} fails when neighboring semiclassical wave packet trajectories approach within a momentum separation of order $l_B^{-1}$.
A finite magnetic field then induces coherent tunneling between the trajectories, as indicated by the red arrows in Figs.~\ref{fig:FS}(a) and (b). 
In the current setup with experimentally relevant range $B\sim0.5 - 4$T, the corresponding scale $a/l_B$ is comparable to the pocket separation over a broad density range, making magnetic breakdown essential in R4G.

\begin{figure}[t]
    \centering
    \includegraphics[width=0.9\linewidth]{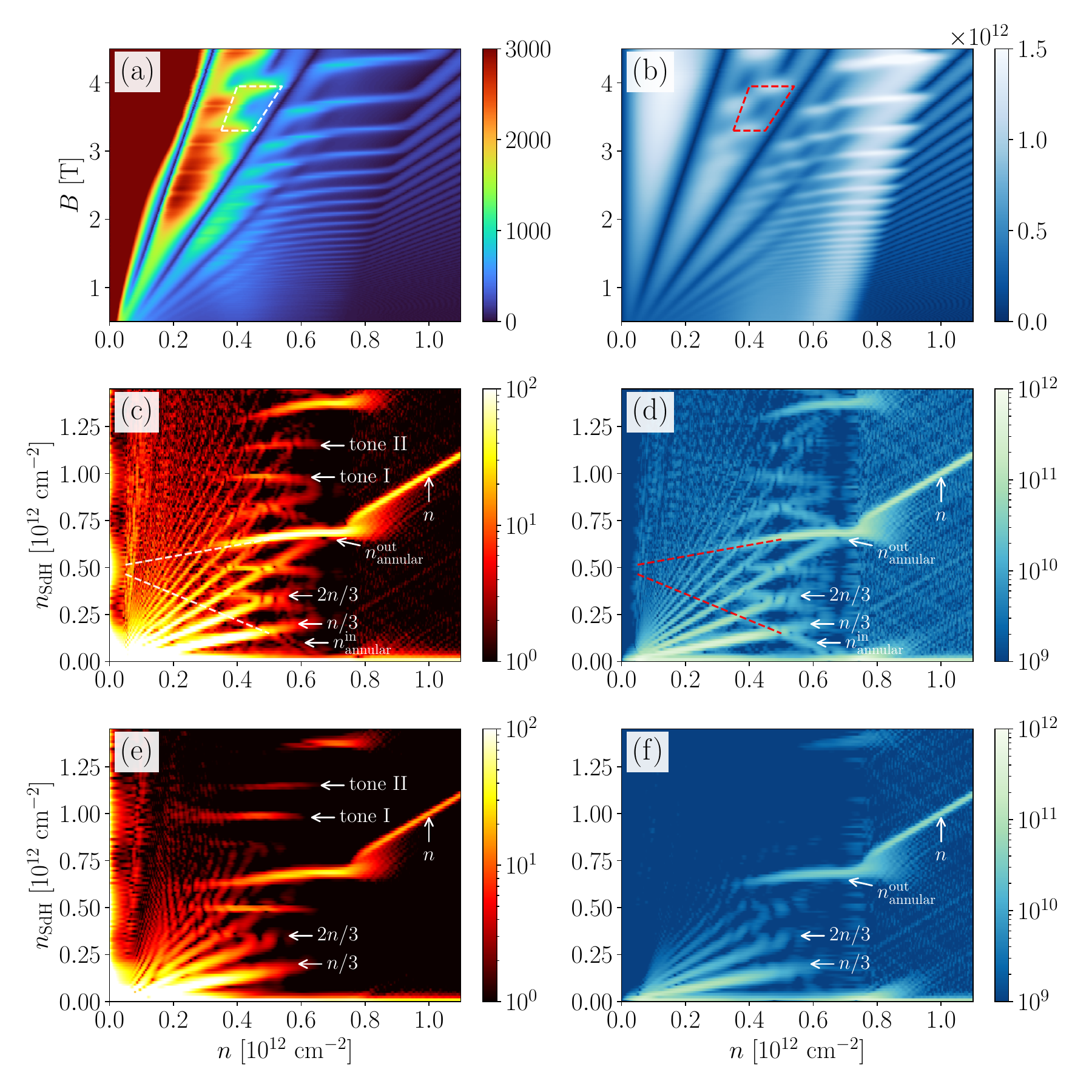}
    \caption{
    Landau-fan plots through both longitudinal resistivity and DOS calculations, together with the corresponding SdH Fourier analyses.
    All calculations are performed at displacement field $u_D=43~\mathrm{meV}$ and disorder level $T_D=0.05~\mathrm{K}$.
    (a) and (b) show the longitudinal resistivity $\rho_{xx}(n, B)$ and DOS $D(n, B)$ at $T=0.5$K, respectively. 
    Examples of the ring-like structures arising from Landau-level crossings are outlined by the white dashed line in (a) and the red dashed line in (b), respectively.
    (c) and (e) show the Fourier spectra of $\rho_{xx}(n, B)$ along the $1/B$ direction over the magnetic field range $B=0.5-8~\mathrm{T}$ at $T=0.2~\mathrm{K}$ and $T=0.5~\mathrm{K}$, respectively. 
    (d) and (f) show the corresponding Fourier spectra of the DOS $D(n, B)$ at $T=0.2~\mathrm{K}$ and $T=0.5~\mathrm{K}$, respectively.
    The white arrows in (c)--(f) indicate the different branches of the SdH frequencies, where ``tone I'' and ``tone II'' denote the two ``multitone'' frequencies, and $n_{\mathrm{annular}}^{\mathrm{in (out)}}$ denotes the carrier densities corresponding to the inner (outer) Fermi surface areas of the annular Fermi sea, respectively.  
    The white and red dashed lines in (c) and (d) show linear extrapolations of the corresponding frequencies to zero density, 
    whose intercepts are one-half of the tone I frequency.
    Note the color scale is logarithmic for better visibility.
    A Hann window is applied before each Fourier transformation to improve the clarity of the Fourier spectra. 
    }
    \label{fig:Resxx}
\end{figure}

To get an intuition of the magnetic breakdown, we perturb the semiclassical description of the three small pockets in Fig.~\ref{fig:FS} (a) with a weak tunneling probability. 
Without magnetic breakdown, the three $C_3$-related pockets produce three degenerate copies of Landau levels, with energy $E_N(B)$ determined from Eq.~\eqref{eqn:OBS}. 
With a tunneling amplitude $t_N(E,B)$ induced by $B$ between neighboring orbits, we can write down an effective Hamiltonian of the Landau levels as 
\begin{equation}\label{eqn:Heff_class}
    H_{\mathrm{eff}}^{\mathrm{MB}} = \begin{pmatrix}
        E_N & -t_N & -t_Ne^{i\Phi_N} \\
        -t_N & E_N & -t_N\\
        -t_Ne^{-i\Phi_N} & -t_N & E_N
    \end{pmatrix}, 
\end{equation}
where $\Phi_N$ is the gauge-invariant phase accumulated 
when the quasi-particle goes around the closed magnetic-breakdown network, 
\begin{equation}
    \Phi_N(E,B) = l_B^2 A_{\mathrm{0}}+\varphi_{N}(E,B)+\mathrm{const.}~,
\end{equation}
where $A_{\mathrm{0}}$ is the leading-order contribution,  
taken to be the extrapolated zero-density intercept of the annular Fermi-surface area. 
$|\varphi_N(E,B)| \ll l_B^2A_{\mathrm{0}}$ contains the remaining phases, including the contributions from the additional Fermi surface area, the tunneling matrix, and the orbital moment and geometric phases. 
We have used the gauge freedom to place the entire phase on one link of Eq.~\eqref{eqn:Heff_class}.
The eigenvalues can be easily solved as 
\begin{equation}\label{eqn:es_MB}
    E_{N,m} = E_N - 2t_N \cos\left(\frac{\Phi_N+2\pi m}{3}\right), \quad m = 0,1,2~.
\end{equation}
More precise derivations of the quantization condition and the form of phase factor $\Phi_N(E,B)$ from a semi-classical model are provided in SM \cite{SM}. 

Magnetic breakdown therefore has two related effects. 
First, it lifts the threefold orbital degeneracy by $t_N$. 
Second, the phase $\Phi_N$ varies approximately periodically with $l_B^2A_{0}\propto 1/B$. 
As a result, the relative energies of the three branches evolve and repeatedly cross each other as the magnetic field is varied [c.f. the lower branches of Fig.~\ref{fig:FS} (e)]. 
Specifically, two of the three branches become degenerate whenever $2\Phi_N=0\ (\mathrm{mod}\ 2\pi)$. 
These crossings generate a characteristic braided structure of the Landau levels and are expected to give rise to a frequency proportional to the area $2A_{\mathrm{0}}$. 
As the Landau-level crossings occur in triplets, the evolution from one triplet to the next further produces a beating envelope, 
resulting in a split, or ``multitone'' structure with a frequency spacing set by the single-pocket Onsager frequency.
We show below that this is indeed the case through SdH Fourier analysis of longitudinal resistivity. 

\textit{Landau Level calculations}---
The experimentally relevant magnetic field strength $0.5\mathrm{T}-4\mathrm{T}$ is too strong for the {perturbatively} semi-classical calculation above.  
Instead, we exactly solve the Landau levels by implementing the Peierls substitution $\bm{k}\rightarrow\bm{\hat{\pi}}$ in the continuum Hamiltonian Eq.~\eqref{eqn:HR4G} as in Refs.~\cite {koshinoTrigonal2009, kalantreFermiology2026}. 
The Hamiltonian is then diagonalized by expressing $\hat{\pi}_\pm \equiv \hat{\pi}_x\pm i\hat{\pi}_y $
in terms of the ladder operators $\pm i\sqrt{2} \hat{a}^\pm/l_B$, 
and truncating the Landau-level basis separately for each layer and sublattice with typical cutoffs of $N_L\approx250$.
This fully quantum calculation automatically incorporates the crossover between isolated pockets and breakdown orbits. 

Fig.~\ref{fig:FS}(e) shows the calculated Landau levels as a function of magnetic field for a given displacement field $u_D = 43$ meV. 
For lower 
energies, the Landau levels originate from the three disconnected electron pockets shown in Fig.~\ref{fig:FS} (a), which are almost degenerate. 
As the magnetic field $B$ increases, the degeneracy is lifted, and the three Landau levels braid with each other to form ring-like structures, consistent with the semiclassical analysis. 
For 
higher energies, the three pockets merge into a single annular Fermi sea. 
The annulus instead supports an outer electron-like orbit with positive slope Landau fans and 
an inner hole-like orbit with negative slope Landau fans. 
The two sets of Landau fans cross each other and also form some ring structures.

\begin{figure}[t]
    \centering
    \includegraphics[width=0.8\linewidth]{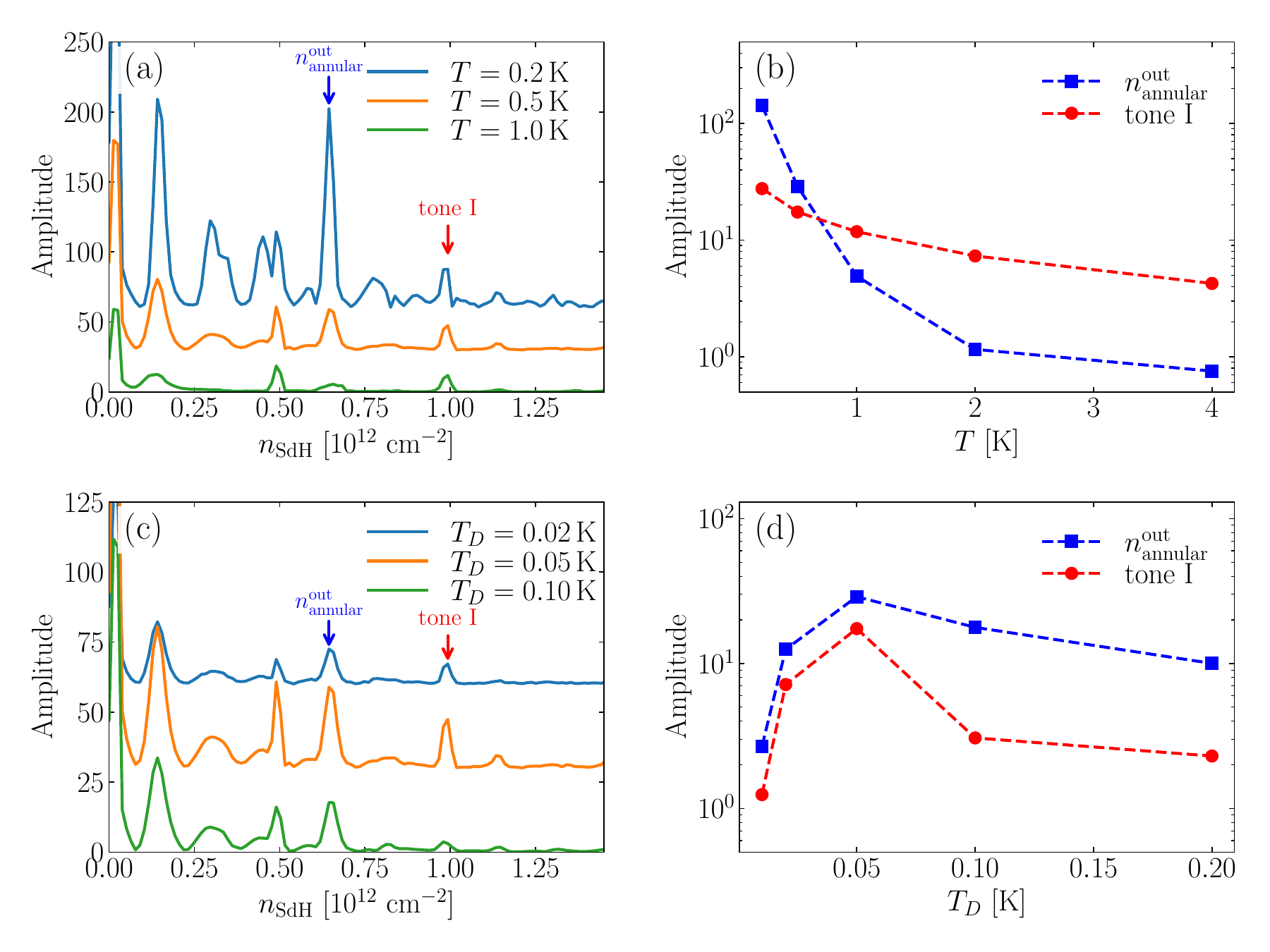}
    \caption{
    (a) Temperature dependence of the SdH frequencies along a line cut with $n=0.45\times 10^{12}\,\mathrm{cm}^{-2}$ in Fig.~\ref{fig:Resxx} (b) for a fixed disorder level corresponding to $T_D=0.05$K. 
    Different temperatures are separated by an offset. 
    The annular Fermi surface and the first tone frequencies are marked by blue and red arrows, respectively. 
    (b) Temperature evolution of the peak value for the annular Fermi surface and first tone peak. 
    (c) Disorder dependence of the SdH frequencies along the same line cut with $n=0.45\times 10^{12}\,\mathrm{cm}^{-2}$. 
    (d) Disorder level evolution of the peak value for the annular Fermi surface and first tone peak. 
    }
    \label{fig:TTD}
\end{figure}

\textit{Quantum oscillations---}
In Fig.~\ref{fig:Resxx}, we show the SdH oscillation in longitudinal resistivity $\rho_{xx}(n,B)$ and its corresponding Fourier spectrum. 
For comparison, we also present the thermally broadened DOS $D(n,B)$ and the corresponding Fourier spectrum of the DOS oscillation. 
\footnote{Strictly speaking, Shubnikov--de Haas oscillations refer to quantum oscillations in longitudinal transport. Throughout this work, we therefore use ``SdH oscillations'' only for the oscillations of $\rho_{xx}$, while the corresponding oscillations in the density of states are referred to as ``DOS oscillations''. }
The Fourier analysis is performed over $B=0.5$--$8~\mathrm{T}$ for every fixed $n$ after applying a Hann window. 
The SdH frequency is expressed as an effective carrier density $n_{\mathrm{SdH}}\equiv ef_{\mathrm{SdH}}/h$, where $f_{\mathrm{SdH}}$ 
is the Fourier frequency conjugate to $1/B$, defined through an oscillatory component of the form $\cos(2\pi f_{\mathrm{SdH}}/B+\phi)$. 
Disorder effects are incorporated through the self-consistent Born approximation \sout{(SCBA)}, with the overall disorder strength characterized by an effective Dingle temperature $T_D$. 
The longitudinal conductivity is evaluated using the Kubo formula at the bubble-diagram level. 
Further details of the transport calculations are provided in the End Matter and SM \cite{SM}.

Three distinct regimes can be identified in the Landau fan map.  
At high densities, $n\gtrsim0.75\times10^{12}\ \mathrm{cm}^{-2}$, the Landau fan originates from the large simply connected Fermi surface as in Fig.~\ref{fig:FS} (c). 
The corresponding SdH oscillation is dominated by the conventional frequency $n_{\mathrm{SdH}}=n$. 
At low densities $n\lesssim 0.5\times10^{12}\ \mathrm{cm}^{-2}$, the Landau fans come from the three nearly degenerate small electron pockets as in Fig.~\ref{fig:FS}(a). 
The corresponding SdH oscillation contains a low-frequency component determined by the area of an individual pocket $n_{\mathrm{SdH}} = n/3$, together with additional higher harmonics $n_{\mathrm{SdH}} = 2n/3, n, \ldots$. 
In the intermediate density regime, the Fermi surface takes an annular shape as in Fig.~\ref{fig:FS} (b),  where the outer electron pocket depends only weakly on the carrier density.  
All these results are consistent with the discussions in Ref.~\cite{kalantreFermiology2026}. 

Importantly, two additional high SdH frequencies are observed as denoted in Figs.~\ref{fig:Resxx}(c) and \ref{fig:Resxx}(e), which match quantitatively well with the experimentally observed ``multitone'' metal~\cite{kalantreFermiology2026}. 
These frequencies remain nearly independent of carrier density over the broad range $n\sim 0.25-0.65 \times10^{12}\,\mathrm{cm}^{-2}$, consistent with the regime governed by the VHSs separating the three-pocket and annular Fermi surfaces.
We note that magnetic breakdown was also considered in Ref.~\cite{kalantreFermiology2026}, although no corresponding anomalous frequencies were identified there. 
This difference originates from the observable considered: whereas the features are weak in the DOS oscillation [Figs.~\ref{fig:Resxx}(d) and \ref{fig:Resxx}(f)], 
these magnetic breakdown structures are considerably more pronounced in the SdH oscillation in longitudinal resistivity $\rho_{xx}$. 
This is because the latter involves products of neighboring spectral functions $A_N(E)A_{N'}(E)$, 
which are strongly enhanced when two Landau levels cross each other. 
Thus, the full transport calculation is necessary, going beyond just DOS considerations. 

These additional peaks originate from the ring-like Landau-level structures generated by repeated crossings and hybridization of the three magnetic breakdown-coupled branches, as indicated by the dashed trapezoid in Figs.~\ref{fig:Resxx}(a) and \ref{fig:Resxx}(b). 
A fixed-density sweep intersects successive rings as the magnetic field is varied, producing an additional oscillation whose frequency is set by the spacing between neighboring rings. 
Because these structures extend over a broad density range, the resulting frequency is nearly density independent. 
Moreover, the Landau levels are organized into successive hybridized triplets inherited from the three symmetry-related pockets. 
The slightly different spacings within and between these triplets produce a beating pattern, which appears as the observed splitting of the high-frequency peak.

\begin{figure}[t]
    \centering
    \includegraphics[width=0.9\linewidth]{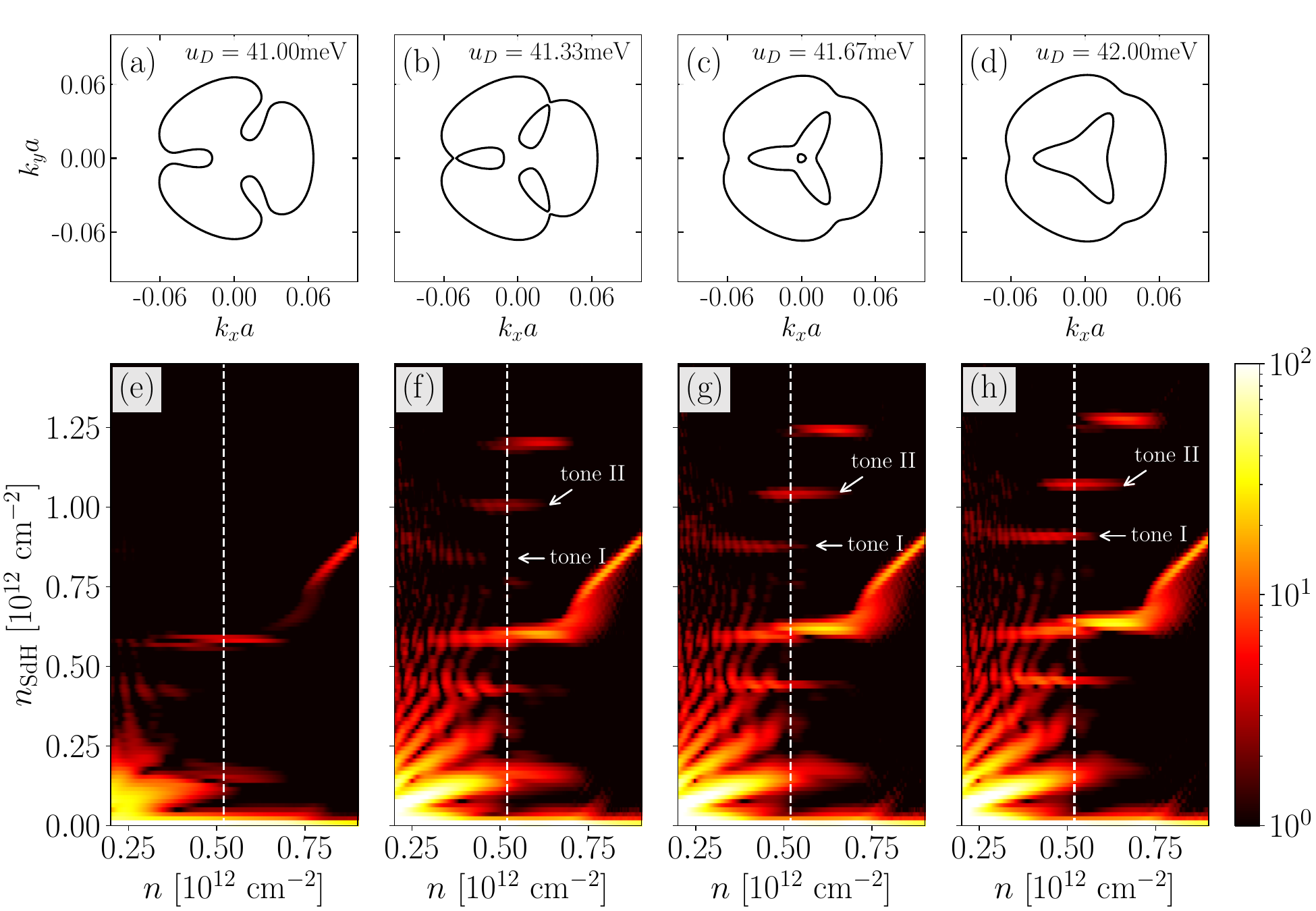}
    \caption{
    SdH oscillation map for displacement near the higher VHSs with $u_D = 41.00, 41.33, 41.67$ and $42.00$ meV, respectively. 
    (a)-(d) Typical Fermi surface geometry at a fixed density 
    $n=0.52\times 10^{12}\,\mathrm{cm}^{-2}$ near the VHS. 
    (e)-(f) The corresponding SdH oscillation plot. 
    All calculations are performed at $T=0.5~\mathrm{K}$ with a disorder broadening corresponding to $T_D=0.05~\mathrm{K}$.
    The ``multitone'' peaks are indicated by the white arrows. 
    The white dashed lines mark the density where the Fermi surface geometry is shown. 
    }
    \label{fig:ud}
\end{figure}

In Fig.~\ref{fig:TTD}, we show the temperature and disorder dependence of the oscillation spectrum along a fixed density line cut along $n_0 = 0.45\times 10^{12}\,\mathrm{cm}^{-2}$. 
With increasing temperature $T$, the SdH oscillations associated with the annular Fermi surface are substantially suppressed. 
By contrast, the magnetic breakdown-induced ``multitone'' peaks remain more robust at high temperature. 
This 
indicates that 
the annular Fermi surface peaks may be thermally suppressed in experiments. 
On the other hand, the ``multitone'' peaks are more sensitive to the disorder broadening of the Landau levels. 
Their weights reach maximal near a disorder level of $T_D\simeq 0.05$K, and decrease at larger $T_D$.  
A moderate level of disorder therefore optimizes the visibility of the magnetic breakdown-induced frequencies. 
More detailed analysis on temperature and disorder can be found in SM \cite{SM}.

\textit{Relation to the CSC---}
Experimentally, the ``multitone'' SdH response expands a noticeable region in the density-displacement map, covering the CSC~\cite{kalantreFermiology2026}. However, it is insufficient to connect the ``multitone'' SdH response to CSC \cite{Ju_private}, especially since these anomalous quantum oscillations mainly come from magnetic breakdown.  
Moreover, a sufficiently large magnetic field may induce significant distortion of Fermi surface, resulting in quantum oscillations which may not directly relate to the zero-field Fermi surface geometry  \cite{Ju_private}. In Ref.~\cite{HanRnGSC2025b}, a finite temperature phase transition manifests in the regime of CSC, suggesting an entropy-driven competitions between two nearby phases \cite{Ju_private}. Thus, a detailed temperature-dependent study is needed for connecting the ``multitone'' features to the normal state of CSC.

On the other hand, previous theories have associated the superconducting instability with the enhanced DOS near the top-right VHSs \cite{chouQM2026, geierCSC2025}. 
In particular, Ref.~\cite{chouQM2026} found that $T_c$ follows the VHS lines and is maximal near the bifurcation of two VHS branches. 
In Fig.~\ref{fig:ud}, we show the SdH oscillations at displacement fields approaching this VHS bifurcation. 
The ``multitone'' feature is most pronounced near the top-left VHS separating the three-pocket and annular Fermi-surface. 
It nevertheless persists at $u_D=41.33$ meV and $u_D=41.67$ meV, where the Fermi-surface geometry is more complicated, and the inner pocket further splits into three small hole pockets. 
By contrast, the feature disappears at $u_D=41.00$ meV, where the three-pocket--annular VHS is absent. 
We also find that the ``multitone'' region overlaps with Fermi surface geometries that support both $C=0$ and $C=1$ CSCs, so it cannot provide evidence about topology \footnote{
For a chiral $p+ip$ state, the BdG Chern-number transition is controlled by whether the chemical potential lies above or below $E(\mathbf{k}=0)$ \cite{chouQM2026,ghoshNonAbelian2010}.
}.  
Finally, the ``multitone'' frequencies increase slightly with $u_D$, consistent with the experiment \cite{kalantreFermiology2026}.

\textit{Conclusion---}
In this letter, we study magnetic breakdown-induced non-Onsager SdH frequencies in rhombohedral tetralayer graphene.  
It provides a natural mechanism for the anomalous quantum oscillations observed experimentally. 
Conversely, our theory also identifies the resulting anomalous oscillations as a direct probe of the underlying VHS. 

The quantitative agreement between our calculation and experiment suggests that the noninteracting band structure already captures the essential Fermi-surface geometry of the parent metallic state of CSC. 
It also suggests the VHS saddle points connecting the three electron pockets remain stable for a wide range of parameter regimes, 
and may play a role in the formation of CSC. 
Since other RnG systems with $n> 4$ have qualitatively similar band structures, we expect these anomalous quantum oscillations to be observed in other graphene multilayers, not just particular to R4G. 

\textit{Acknowledgments--} 
We thank Long Ju for helpful discussions and for sharing unpublished experimental data with us. 
We thank Daniel E. Parker for pointing out a correction to the Onsager-Bohr-Sommerfeld condition. This work is supported by the Laboratory for Physical Sciences through its support of the Condensed Matter Theory Center at the University of Maryland. 

\textit{Data availability--} The data that support the findings of this article are not publicly available. The data are available from the authors upon reasonable request.

\bibliography{refs}

\begin{thebibliography}{49}%
\makeatletter
\providecommand \@ifxundefined [1]{%
 \@ifx{#1\undefined}
}%
\providecommand \@ifnum [1]{%
 \ifnum #1\expandafter \@firstoftwo
 \else \expandafter \@secondoftwo
 \fi
}%
\providecommand \@ifx [1]{%
 \ifx #1\expandafter \@firstoftwo
 \else \expandafter \@secondoftwo
 \fi
}%
\providecommand \natexlab [1]{#1}%
\providecommand \enquote  [1]{``#1''}%
\providecommand \bibnamefont  [1]{#1}%
\providecommand \bibfnamefont [1]{#1}%
\providecommand \citenamefont [1]{#1}%
\providecommand \href@noop [0]{\@secondoftwo}%
\providecommand \href [0]{\begingroup \@sanitize@url \@href}%
\providecommand \@href[1]{\@@startlink{#1}\@@href}%
\providecommand \@@href[1]{\endgroup#1\@@endlink}%
\providecommand \@sanitize@url [0]{\catcode `\\12\catcode `\$12\catcode
  `\&12\catcode `\#12\catcode `\^12\catcode `\_12\catcode `\%12\relax}%
\providecommand \@@startlink[1]{}%
\providecommand \@@endlink[0]{}%
\providecommand \url  [0]{\begingroup\@sanitize@url \@url }%
\providecommand \@url [1]{\endgroup\@href {#1}{\urlprefix }}%
\providecommand \urlprefix  [0]{URL }%
\providecommand \Eprint [0]{\href }%
\providecommand \doibase [0]{https://doi.org/}%
\providecommand \selectlanguage [0]{\@gobble}%
\providecommand \bibinfo  [0]{\@secondoftwo}%
\providecommand \bibfield  [0]{\@secondoftwo}%
\providecommand \translation [1]{[#1]}%
\providecommand \BibitemOpen [0]{}%
\providecommand \bibitemStop [0]{}%
\providecommand \bibitemNoStop [0]{.\EOS\space}%
\providecommand \EOS [0]{\spacefactor3000\relax}%
\providecommand \BibitemShut  [1]{\csname bibitem#1\endcsname}%
\let\auto@bib@innerbib\@empty
\bibitem [{\citenamefont {Han}\ \emph {et~al.}(2025)\citenamefont {Han},
  \citenamefont {Lu}, \citenamefont {Hadjri}, \citenamefont {Shi},
  \citenamefont {Wu}, \citenamefont {Xu}, \citenamefont {Yao}, \citenamefont
  {Cotten}, \citenamefont {Sharifi~Sedeh}, \citenamefont {Weldeyesus},
  \citenamefont {Yang}, \citenamefont {Seo}, \citenamefont {Ye}, \citenamefont
  {Zhou}, \citenamefont {Liu}, \citenamefont {Shi}, \citenamefont {Hua},
  \citenamefont {Watanabe}, \citenamefont {Taniguchi}, \citenamefont {Xiong},
  \citenamefont {Zumb{\"u}hl}, \citenamefont {Fu},\ and\ \citenamefont
  {Ju}}]{HanRnGSC2025b}%
  \BibitemOpen
  \bibfield  {author} {\bibinfo {author} {\bibfnamefont {T.}~\bibnamefont
  {Han}}, \bibinfo {author} {\bibfnamefont {Z.}~\bibnamefont {Lu}}, \bibinfo
  {author} {\bibfnamefont {Z.}~\bibnamefont {Hadjri}}, \bibinfo {author}
  {\bibfnamefont {L.}~\bibnamefont {Shi}}, \bibinfo {author} {\bibfnamefont
  {Z.}~\bibnamefont {Wu}}, \bibinfo {author} {\bibfnamefont {W.}~\bibnamefont
  {Xu}}, \bibinfo {author} {\bibfnamefont {Y.}~\bibnamefont {Yao}}, \bibinfo
  {author} {\bibfnamefont {A.~A.}\ \bibnamefont {Cotten}}, \bibinfo {author}
  {\bibfnamefont {O.}~\bibnamefont {Sharifi~Sedeh}}, \bibinfo {author}
  {\bibfnamefont {H.}~\bibnamefont {Weldeyesus}}, \bibinfo {author}
  {\bibfnamefont {J.}~\bibnamefont {Yang}}, \bibinfo {author} {\bibfnamefont
  {J.}~\bibnamefont {Seo}}, \bibinfo {author} {\bibfnamefont {S.}~\bibnamefont
  {Ye}}, \bibinfo {author} {\bibfnamefont {M.}~\bibnamefont {Zhou}}, \bibinfo
  {author} {\bibfnamefont {H.}~\bibnamefont {Liu}}, \bibinfo {author}
  {\bibfnamefont {G.}~\bibnamefont {Shi}}, \bibinfo {author} {\bibfnamefont
  {Z.}~\bibnamefont {Hua}}, \bibinfo {author} {\bibfnamefont {K.}~\bibnamefont
  {Watanabe}}, \bibinfo {author} {\bibfnamefont {T.}~\bibnamefont {Taniguchi}},
  \bibinfo {author} {\bibfnamefont {P.}~\bibnamefont {Xiong}}, \bibinfo
  {author} {\bibfnamefont {D.~M.}\ \bibnamefont {Zumb{\"u}hl}}, \bibinfo
  {author} {\bibfnamefont {L.}~\bibnamefont {Fu}},\ and\ \bibinfo {author}
  {\bibfnamefont {L.}~\bibnamefont {Ju}},\ }\bibfield  {title} {\bibinfo
  {title} {Signatures of chiral superconductivity in rhombohedral graphene},\
  }\href {https://doi.org/10.1038/s41586-025-09169-7} {\bibfield  {journal}
  {\bibinfo  {journal} {Nature}\ }\textbf {\bibinfo {volume} {643}},\ \bibinfo
  {pages} {654} (\bibinfo {year} {2025})}\BibitemShut {NoStop}%
\bibitem [{\citenamefont {Morissette}\ \emph {et~al.}(2025)\citenamefont
  {Morissette}, \citenamefont {Qin}, \citenamefont {Wu}, \citenamefont
  {Watanabe}, \citenamefont {Taniguchi},\ and\ \citenamefont {Li}}]{JIAli2025}%
  \BibitemOpen
  \bibfield  {author} {\bibinfo {author} {\bibfnamefont {E.}~\bibnamefont
  {Morissette}}, \bibinfo {author} {\bibfnamefont {P.}~\bibnamefont {Qin}},
  \bibinfo {author} {\bibfnamefont {H.~T.}\ \bibnamefont {Wu}}, \bibinfo
  {author} {\bibfnamefont {K.}~\bibnamefont {Watanabe}}, \bibinfo {author}
  {\bibfnamefont {T.}~\bibnamefont {Taniguchi}},\ and\ \bibinfo {author}
  {\bibfnamefont {J.~I.~A.}\ \bibnamefont {Li}},\ }\href@noop {} {\bibinfo
  {title} {Superconductivity, {{Anomalous Hall Effect}}, and {{Stripe Order}}
  in {{Rhombohedral Hexalayer Graphene}}}} (\bibinfo {year} {2025}),\ \Eprint
  {https://arxiv.org/abs/2504.05129} {arXiv:2504.05129} \BibitemShut {NoStop}%
\bibitem [{\citenamefont {Kalantre}\ \emph {et~al.}(2026)\citenamefont
  {Kalantre}, \citenamefont {Alexander}, \citenamefont {{May-Mann}},
  \citenamefont {{Herzog-Arbeitman}}, \citenamefont {Hocking}, \citenamefont
  {Cao}, \citenamefont {Watanabe}, \citenamefont {Taniguchi}, \citenamefont
  {{Goldhaber-Gordon}}, \citenamefont {Mannix}, \citenamefont {Devakul},
  \citenamefont {Kwan}, \citenamefont {Parker},\ and\ \citenamefont
  {Sharpe}}]{kalantreFermiology2026}%
  \BibitemOpen
  \bibfield  {author} {\bibinfo {author} {\bibfnamefont {S.~S.}\ \bibnamefont
  {Kalantre}}, \bibinfo {author} {\bibfnamefont {B.~H.}\ \bibnamefont
  {Alexander}}, \bibinfo {author} {\bibfnamefont {J.}~\bibnamefont
  {{May-Mann}}}, \bibinfo {author} {\bibfnamefont {J.}~\bibnamefont
  {{Herzog-Arbeitman}}}, \bibinfo {author} {\bibfnamefont {M.}~\bibnamefont
  {Hocking}}, \bibinfo {author} {\bibfnamefont {Q.}~\bibnamefont {Cao}},
  \bibinfo {author} {\bibfnamefont {K.}~\bibnamefont {Watanabe}}, \bibinfo
  {author} {\bibfnamefont {T.}~\bibnamefont {Taniguchi}}, \bibinfo {author}
  {\bibfnamefont {D.}~\bibnamefont {{Goldhaber-Gordon}}}, \bibinfo {author}
  {\bibfnamefont {A.~J.}\ \bibnamefont {Mannix}}, \bibinfo {author}
  {\bibfnamefont {T.}~\bibnamefont {Devakul}}, \bibinfo {author} {\bibfnamefont
  {Y.~H.}\ \bibnamefont {Kwan}}, \bibinfo {author} {\bibfnamefont {D.~E.}\
  \bibnamefont {Parker}},\ and\ \bibinfo {author} {\bibfnamefont
  {A.}~\bibnamefont {Sharpe}},\ }\href@noop {} {\bibinfo {title} {Fermiology
  and the {{Candidate Chiral Superconductor}} in {{Rhombohedral Tetralayer
  Graphene}}}} (\bibinfo {year} {2026}),\ \Eprint
  {https://arxiv.org/abs/2606.05356} {arXiv:2606.05356} \BibitemShut {NoStop}%
\bibitem [{\citenamefont {Hua}\ \emph {et~al.}(2026)\citenamefont {Hua},
  \citenamefont {Ye}, \citenamefont {Pattanakanvijit}, \citenamefont {Shi},
  \citenamefont {Han}, \citenamefont {Aitken}, \citenamefont {Yang},
  \citenamefont {Seo}, \citenamefont {Liu}, \citenamefont {Hao}, \citenamefont
  {Xiao}, \citenamefont {Guo}, \citenamefont {Phong}, \citenamefont {Watanabe},
  \citenamefont {Taniguchi}, \citenamefont {Huang}, \citenamefont
  {Lewandowski}, \citenamefont {Ju}, \citenamefont {Xiong},\ and\ \citenamefont
  {Lu}}]{huaR6G2026}%
  \BibitemOpen
  \bibfield  {author} {\bibinfo {author} {\bibfnamefont {Z.}~\bibnamefont
  {Hua}}, \bibinfo {author} {\bibfnamefont {S.}~\bibnamefont {Ye}}, \bibinfo
  {author} {\bibfnamefont {P.}~\bibnamefont {Pattanakanvijit}}, \bibinfo
  {author} {\bibfnamefont {G.}~\bibnamefont {Shi}}, \bibinfo {author}
  {\bibfnamefont {T.}~\bibnamefont {Han}}, \bibinfo {author} {\bibfnamefont
  {E.}~\bibnamefont {Aitken}}, \bibinfo {author} {\bibfnamefont
  {J.}~\bibnamefont {Yang}}, \bibinfo {author} {\bibfnamefont {J.}~\bibnamefont
  {Seo}}, \bibinfo {author} {\bibfnamefont {H.}~\bibnamefont {Liu}}, \bibinfo
  {author} {\bibfnamefont {R.}~\bibnamefont {Hao}}, \bibinfo {author}
  {\bibfnamefont {K.}~\bibnamefont {Xiao}}, \bibinfo {author} {\bibfnamefont
  {J.}~\bibnamefont {Guo}}, \bibinfo {author} {\bibfnamefont {V.~T.}\
  \bibnamefont {Phong}}, \bibinfo {author} {\bibfnamefont {K.}~\bibnamefont
  {Watanabe}}, \bibinfo {author} {\bibfnamefont {T.}~\bibnamefont {Taniguchi}},
  \bibinfo {author} {\bibfnamefont {C.}~\bibnamefont {Huang}}, \bibinfo
  {author} {\bibfnamefont {C.}~\bibnamefont {Lewandowski}}, \bibinfo {author}
  {\bibfnamefont {L.}~\bibnamefont {Ju}}, \bibinfo {author} {\bibfnamefont
  {P.}~\bibnamefont {Xiong}},\ and\ \bibinfo {author} {\bibfnamefont
  {Z.}~\bibnamefont {Lu}},\ }\href@noop {} {\bibinfo {title} {Multi-{{Knob
  Switchable Chiral Superconductivity Quartet}} in {{Rhombohedral Graphene}}}}
  (\bibinfo {year} {2026}),\ \Eprint {https://arxiv.org/abs/2607.06520}
  {arXiv:2607.06520} \BibitemShut {NoStop}%
\bibitem [{\citenamefont {Geier}\ \emph {et~al.}(2025)\citenamefont {Geier},
  \citenamefont {Davydova},\ and\ \citenamefont {Fu}}]{geierCSC2025}%
  \BibitemOpen
  \bibfield  {author} {\bibinfo {author} {\bibfnamefont {M.}~\bibnamefont
  {Geier}}, \bibinfo {author} {\bibfnamefont {M.}~\bibnamefont {Davydova}},\
  and\ \bibinfo {author} {\bibfnamefont {L.}~\bibnamefont {Fu}},\ }\bibfield
  {title} {\bibinfo {title} {Chiral and topological superconductivity in
  isospin polarized multilayer graphene},\ }\href
  {https://doi.org/10.1038/s41467-025-66902-6} {\bibfield  {journal} {\bibinfo
  {journal} {Nature Communications}\ }\textbf {\bibinfo {volume} {17}},\
  \bibinfo {pages} {232} (\bibinfo {year} {2025})}\BibitemShut {NoStop}%
\bibitem [{\citenamefont {Chou}\ \emph {et~al.}(2025)\citenamefont {Chou},
  \citenamefont {Zhu},\ and\ \citenamefont {Das~Sarma}}]{chouR4GSC2025}%
  \BibitemOpen
  \bibfield  {author} {\bibinfo {author} {\bibfnamefont {Y.-Z.}\ \bibnamefont
  {Chou}}, \bibinfo {author} {\bibfnamefont {J.}~\bibnamefont {Zhu}},\ and\
  \bibinfo {author} {\bibfnamefont {S.}~\bibnamefont {Das~Sarma}},\ }\bibfield
  {title} {\bibinfo {title} {Intravalley spin-polarized superconductivity in
  rhombohedral tetralayer graphene},\ }\href
  {https://doi.org/10.1103/PhysRevB.111.174523} {\bibfield  {journal} {\bibinfo
   {journal} {Physical Review B}\ }\textbf {\bibinfo {volume} {111}},\ \bibinfo
  {pages} {174523} (\bibinfo {year} {2025})}\BibitemShut {NoStop}%
\bibitem [{\citenamefont {Yang}\ and\ \citenamefont
  {Zhang}(2025)}]{yangPDW2025}%
  \BibitemOpen
  \bibfield  {author} {\bibinfo {author} {\bibfnamefont {H.}~\bibnamefont
  {Yang}}\ and\ \bibinfo {author} {\bibfnamefont {Y.-H.}\ \bibnamefont
  {Zhang}},\ }\bibfield  {title} {\bibinfo {title} {Topological incommensurate
  {{Fulde-Ferrell-Larkin-Ovchinnikov}} superconductor and {{Bogoliubov Fermi}}
  surface in rhombohedral tetralayer graphene},\ }\href
  {https://doi.org/10.1103/k8s3-dgfs} {\bibfield  {journal} {\bibinfo
  {journal} {Physical Review B}\ }\textbf {\bibinfo {volume} {112}},\ \bibinfo
  {pages} {L020506} (\bibinfo {year} {2025})}\BibitemShut {NoStop}%
\bibitem [{\citenamefont {{Parra-Mart{\'i}nez}}\ \emph
  {et~al.}(2025)\citenamefont {{Parra-Mart{\'i}nez}}, \citenamefont
  {{Jimeno-Pozo}}, \citenamefont {Phong}, \citenamefont {{Sainz-Cruz}},
  \citenamefont {Kaplan}, \citenamefont {Emanuel}, \citenamefont {Oreg},
  \citenamefont {Pantale{\'o}n}, \citenamefont {{Silva-Guill{\'e}n}},\ and\
  \citenamefont {Guinea}}]{parraRPA2025}%
  \BibitemOpen
  \bibfield  {author} {\bibinfo {author} {\bibfnamefont {G.}~\bibnamefont
  {{Parra-Mart{\'i}nez}}}, \bibinfo {author} {\bibfnamefont {A.}~\bibnamefont
  {{Jimeno-Pozo}}}, \bibinfo {author} {\bibfnamefont {V.~T.}\ \bibnamefont
  {Phong}}, \bibinfo {author} {\bibfnamefont {H.}~\bibnamefont {{Sainz-Cruz}}},
  \bibinfo {author} {\bibfnamefont {D.}~\bibnamefont {Kaplan}}, \bibinfo
  {author} {\bibfnamefont {P.}~\bibnamefont {Emanuel}}, \bibinfo {author}
  {\bibfnamefont {Y.}~\bibnamefont {Oreg}}, \bibinfo {author} {\bibfnamefont
  {P.~A.}\ \bibnamefont {Pantale{\'o}n}}, \bibinfo {author} {\bibfnamefont
  {J.~{\'A}.}\ \bibnamefont {{Silva-Guill{\'e}n}}},\ and\ \bibinfo {author}
  {\bibfnamefont {F.}~\bibnamefont {Guinea}},\ }\bibfield  {title} {\bibinfo
  {title} {Band {{Renormalization}}, {{Quarter Metals}}, and {{Chiral
  Superconductivity}} in {{Rhombohedral Tetralayer Graphene}}},\ }\href
  {https://doi.org/10.1103/zfmh-rjzc} {\bibfield  {journal} {\bibinfo
  {journal} {Physical Review Letters}\ }\textbf {\bibinfo {volume} {135}},\
  \bibinfo {pages} {136503} (\bibinfo {year} {2025})}\BibitemShut {NoStop}%
\bibitem [{\citenamefont {Gaggioli}\ \emph {et~al.}(2025)\citenamefont
  {Gaggioli}, \citenamefont {Guerci},\ and\ \citenamefont
  {Fu}}]{gaggioliVAV2025}%
  \BibitemOpen
  \bibfield  {author} {\bibinfo {author} {\bibfnamefont {F.}~\bibnamefont
  {Gaggioli}}, \bibinfo {author} {\bibfnamefont {D.}~\bibnamefont {Guerci}},\
  and\ \bibinfo {author} {\bibfnamefont {L.}~\bibnamefont {Fu}},\ }\bibfield
  {title} {\bibinfo {title} {Spontaneous {{Vortex-Antivortex Lattice}} and
  {{Majorana Fermions}} in {{Rhombohedral Graphene}}},\ }\href
  {https://doi.org/10.1103/k8sb-rqxf} {\bibfield  {journal} {\bibinfo
  {journal} {Physical Review Letters}\ }\textbf {\bibinfo {volume} {135}},\
  \bibinfo {pages} {116001} (\bibinfo {year} {2025})}\BibitemShut {NoStop}%
\bibitem [{\citenamefont {Chou}(2026)}]{chouQM2026}%
  \BibitemOpen
  \bibfield  {author} {\bibinfo {author} {\bibfnamefont {Y.-Z.}\ \bibnamefont
  {Chou}},\ }\bibfield  {title} {\bibinfo {title} {Superconductivity from
  {{Phonon-Mediated Retardation}} in a {{Single-Flavor Metal}}},\ }\bibfield
  {journal} {\bibinfo  {journal} {Physical Review Letters}\ }\textbf {\bibinfo
  {volume} {137}},\ \href {https://doi.org/10.1103/dcn2-lc8k}
  {10.1103/dcn2-lc8k} (\bibinfo {year} {2026})\BibitemShut {NoStop}%
\bibitem [{\citenamefont {Christos}\ \emph {et~al.}(2026)\citenamefont
  {Christos}, \citenamefont {Bonetti},\ and\ \citenamefont
  {Scheurer}}]{christosPDW2025}%
  \BibitemOpen
  \bibfield  {author} {\bibinfo {author} {\bibfnamefont {M.}~\bibnamefont
  {Christos}}, \bibinfo {author} {\bibfnamefont {P.~M.}\ \bibnamefont
  {Bonetti}},\ and\ \bibinfo {author} {\bibfnamefont {M.~S.}\ \bibnamefont
  {Scheurer}},\ }\bibfield  {title} {\bibinfo {title} {Finite-{{Momentum
  Pairing}} and {{Superlattice Superconductivity}} in {{Valley-Imbalanced
  Rhombohedral Graphene}}},\ }\href {https://doi.org/10.1103/7xj7-hn98}
  {\bibfield  {journal} {\bibinfo  {journal} {Phys. Rev. Lett.}\ }\textbf
  {\bibinfo {volume} {137}},\ \bibinfo {pages} {046505} (\bibinfo {year}
  {2026})}\BibitemShut {NoStop}%
\bibitem [{\citenamefont {Sau}\ and\ \citenamefont
  {Wang}(2024)}]{sauAnomalousHall2024}%
  \BibitemOpen
  \bibfield  {author} {\bibinfo {author} {\bibfnamefont {J.~D.}\ \bibnamefont
  {Sau}}\ and\ \bibinfo {author} {\bibfnamefont {S.}~\bibnamefont {Wang}},\
  }\href@noop {} {\bibinfo {title} {Theory of anomalous {{Hall}} effect from
  screened vortex charge in a phase disordered superconductor}} (\bibinfo
  {year} {2024}),\ \Eprint {https://arxiv.org/abs/2411.08969}
  {arXiv:2411.08969} \BibitemShut {NoStop}%
\bibitem [{\citenamefont {Zhu}\ and\ \citenamefont {Huang}(2026)}]{zhuCSC2026}%
  \BibitemOpen
  \bibfield  {author} {\bibinfo {author} {\bibfnamefont {J.}~\bibnamefont
  {Zhu}}\ and\ \bibinfo {author} {\bibfnamefont {C.}~\bibnamefont {Huang}},\
  }\href@noop {} {\bibinfo {title} {Microscopic origin of orbital magnetization
  in chiral superconductors}} (\bibinfo {year} {2026}),\ \Eprint
  {https://arxiv.org/abs/2601.12387} {arXiv:2601.12387} \BibitemShut {NoStop}%
\bibitem [{\citenamefont {Gil}\ and\ \citenamefont {Berg}(2026)}]{gilPDW2026}%
  \BibitemOpen
  \bibfield  {author} {\bibinfo {author} {\bibfnamefont {A.}~\bibnamefont
  {Gil}}\ and\ \bibinfo {author} {\bibfnamefont {E.}~\bibnamefont {Berg}},\
  }\bibfield  {title} {\bibinfo {title} {Charge and pair density waves in a
  spin- and valley-polarized system at a {{Van Hove}} singularity},\ }\href
  {https://doi.org/10.1103/b39r-bjxs} {\bibfield  {journal} {\bibinfo
  {journal} {Physical Review B}\ }\textbf {\bibinfo {volume} {113}},\ \bibinfo
  {pages} {075154} (\bibinfo {year} {2026})}\BibitemShut {NoStop}%
\bibitem [{\citenamefont {Dong}\ and\ \citenamefont {Lee}(2026)}]{dongQM2026}%
  \BibitemOpen
  \bibfield  {author} {\bibinfo {author} {\bibfnamefont {Z.}~\bibnamefont
  {Dong}}\ and\ \bibinfo {author} {\bibfnamefont {P.~A.}\ \bibnamefont {Lee}},\
  }\bibfield  {title} {\bibinfo {title} {Controlled expansion for pairing in a
  polarized band with strong repulsion},\ }\href
  {https://doi.org/10.1103/yw2c-qyj1} {\bibfield  {journal} {\bibinfo
  {journal} {Physical Review B}\ }\textbf {\bibinfo {volume} {113}},\ \bibinfo
  {pages} {104501} (\bibinfo {year} {2026})}\BibitemShut {NoStop}%
\bibitem [{\citenamefont {Qin}\ and\ \citenamefont {Wu}(2026)}]{qinR4G2026}%
  \BibitemOpen
  \bibfield  {author} {\bibinfo {author} {\bibfnamefont {Q.}~\bibnamefont
  {Qin}}\ and\ \bibinfo {author} {\bibfnamefont {C.}~\bibnamefont {Wu}},\
  }\bibfield  {title} {\bibinfo {title} {Chiral {{Finite-Momentum
  Superconductivity}} in the {{Tetralayer Graphene}}},\ }\href
  {https://doi.org/10.1088/0256-307X/43/3/030708} {\bibfield  {journal}
  {\bibinfo  {journal} {Chinese Physics Letters}\ }\textbf {\bibinfo {volume}
  {43}},\ \bibinfo {pages} {030708} (\bibinfo {year} {2026})}\BibitemShut
  {NoStop}%
\bibitem [{\citenamefont {Chen}\ \emph {et~al.}(2025)\citenamefont {Chen},
  \citenamefont {Scheurer},\ and\ \citenamefont {Schrade}}]{chenDiode2025}%
  \BibitemOpen
  \bibfield  {author} {\bibinfo {author} {\bibfnamefont {Y.}~\bibnamefont
  {Chen}}, \bibinfo {author} {\bibfnamefont {M.~S.}\ \bibnamefont {Scheurer}},\
  and\ \bibinfo {author} {\bibfnamefont {C.}~\bibnamefont {Schrade}},\
  }\bibfield  {title} {\bibinfo {title} {Intrinsic superconducting diode effect
  and nonreciprocal superconductivity in rhombohedral graphene multilayers},\
  }\href {https://doi.org/10.1103/zgnk-rw1p} {\bibfield  {journal} {\bibinfo
  {journal} {Physical Review B}\ }\textbf {\bibinfo {volume} {112}},\ \bibinfo
  {pages} {L060505} (\bibinfo {year} {2025})}\BibitemShut {NoStop}%
\bibitem [{\citenamefont {Jahin}\ and\ \citenamefont
  {Lin}(2026)}]{jahinKohnLuttinger2026}%
  \BibitemOpen
  \bibfield  {author} {\bibinfo {author} {\bibfnamefont {A.}~\bibnamefont
  {Jahin}}\ and\ \bibinfo {author} {\bibfnamefont {S.-Z.}\ \bibnamefont
  {Lin}},\ }\bibfield  {title} {\bibinfo {title} {Enhanced {{Kohn-Luttinger}}
  superconductivity in geometric bands},\ }\href
  {https://doi.org/10.1103/gt8h-czf3} {\bibfield  {journal} {\bibinfo
  {journal} {Physical Review B}\ }\textbf {\bibinfo {volume} {113}},\ \bibinfo
  {pages} {014504} (\bibinfo {year} {2026})}\BibitemShut {NoStop}%
\bibitem [{\citenamefont {Li}\ \emph {et~al.}(2025)\citenamefont {Li},
  \citenamefont {Kwan}, \citenamefont {Yao},\ and\ \citenamefont
  {Bernevig}}]{liBerry2025}%
  \BibitemOpen
  \bibfield  {author} {\bibinfo {author} {\bibfnamefont {M.-R.}\ \bibnamefont
  {Li}}, \bibinfo {author} {\bibfnamefont {Y.~H.}\ \bibnamefont {Kwan}},
  \bibinfo {author} {\bibfnamefont {H.}~\bibnamefont {Yao}},\ and\ \bibinfo
  {author} {\bibfnamefont {B.~A.}\ \bibnamefont {Bernevig}},\ }\href@noop {}
  {\bibinfo {title} {Berry {{Trashcan With Short Range Attraction}}:{{Exact}}
  $p_x+i p_y$ {{Superconductivity}} in {{Rhombohedral Graphene}}}} (\bibinfo
  {year} {2025}),\ \Eprint {https://arxiv.org/abs/2509.16312}
  {arXiv:2509.16312} \BibitemShut {NoStop}%
\bibitem [{\citenamefont {{May-Mann}}\ \emph {et~al.}(2026)\citenamefont
  {{May-Mann}}, \citenamefont {Helbig},\ and\ \citenamefont
  {Devakul}}]{maymannBerry2026}%
  \BibitemOpen
  \bibfield  {author} {\bibinfo {author} {\bibfnamefont {J.}~\bibnamefont
  {{May-Mann}}}, \bibinfo {author} {\bibfnamefont {T.}~\bibnamefont {Helbig}},\
  and\ \bibinfo {author} {\bibfnamefont {T.}~\bibnamefont {Devakul}},\
  }\bibfield  {title} {\bibinfo {title} {How pairing mechanism dictates
  topology in valley-polarized superconductors with {{Berry}} curvature},\
  }\href {https://doi.org/10.1038/s41535-026-00878-4} {\bibfield  {journal}
  {\bibinfo  {journal} {npj Quantum Materials}\ }\textbf {\bibinfo {volume}
  {11}},\ \bibinfo {pages} {53} (\bibinfo {year} {2026})}\BibitemShut {NoStop}%
\bibitem [{\citenamefont {Patri}\ and\ \citenamefont
  {Franz}(2025)}]{patriRnG2025}%
  \BibitemOpen
  \bibfield  {author} {\bibinfo {author} {\bibfnamefont {A.~S.}\ \bibnamefont
  {Patri}}\ and\ \bibinfo {author} {\bibfnamefont {M.}~\bibnamefont {Franz}},\
  }\bibfield  {title} {\bibinfo {title} {Family of multilayer graphene
  superconductors with tunable chirality: {{Momentum-space}} vortices nucleated
  by a ring of {{Berry}} curvature},\ }\href
  {https://doi.org/10.1103/pgqh-wm5v} {\bibfield  {journal} {\bibinfo
  {journal} {Physical Review B}\ }\textbf {\bibinfo {volume} {112}},\ \bibinfo
  {pages} {214505} (\bibinfo {year} {2025})}\BibitemShut {NoStop}%
\bibitem [{\citenamefont {Wang}\ \emph {et~al.}(2026)\citenamefont {Wang},
  \citenamefont {Gao},\ and\ \citenamefont {Yang}}]{wangCSC2026}%
  \BibitemOpen
  \bibfield  {author} {\bibinfo {author} {\bibfnamefont {Y.-Q.}\ \bibnamefont
  {Wang}}, \bibinfo {author} {\bibfnamefont {Z.-Q.}\ \bibnamefont {Gao}},\ and\
  \bibinfo {author} {\bibfnamefont {H.}~\bibnamefont {Yang}},\ }\bibfield
  {title} {\bibinfo {title} {Chiral superconductivity from a parent {{Chern}}
  band and its non-{{Abelian}} generalization},\ }\href
  {https://doi.org/10.1103/fdz1-dbf6} {\bibfield  {journal} {\bibinfo
  {journal} {Physical Review B}\ }\textbf {\bibinfo {volume} {113}},\ \bibinfo
  {pages} {174507} (\bibinfo {year} {2026})}\BibitemShut {NoStop}%
\bibitem [{\citenamefont {Yoon}\ \emph {et~al.}(2026)\citenamefont {Yoon},
  \citenamefont {Xu}, \citenamefont {Barlas},\ and\ \citenamefont
  {Zhang}}]{yoonQMl2026a}%
  \BibitemOpen
  \bibfield  {author} {\bibinfo {author} {\bibfnamefont {C.}~\bibnamefont
  {Yoon}}, \bibinfo {author} {\bibfnamefont {T.}~\bibnamefont {Xu}}, \bibinfo
  {author} {\bibfnamefont {Y.}~\bibnamefont {Barlas}},\ and\ \bibinfo {author}
  {\bibfnamefont {F.}~\bibnamefont {Zhang}},\ }\bibfield  {title} {\bibinfo
  {title} {Quarter-{{Metal Superconductivity}} in {{Rhombohedral Graphene}}},\
  }\href {https://doi.org/10.1103/fcdc-9lm3} {\bibfield  {journal} {\bibinfo
  {journal} {Physical Review Letters}\ }\textbf {\bibinfo {volume} {136}},\
  \bibinfo {pages} {026603} (\bibinfo {year} {2026})}\BibitemShut {NoStop}%
\bibitem [{\citenamefont {Pippard}(1962)}]{pippard1962a}%
  \BibitemOpen
  \bibfield  {author} {\bibinfo {author} {\bibfnamefont {A.~B.}\ \bibnamefont
  {Pippard}},\ }\bibfield  {title} {\bibinfo {title} {Quantization of coupled
  orbits in metals},\ }\href {https://doi.org/10.1098/rspa.1962.0200}
  {\bibfield  {journal} {\bibinfo  {journal} {Proceedings of the Royal Society
  of London. A. Mathematical and Physical Sciences}\ }\textbf {\bibinfo
  {volume} {270}},\ \bibinfo {pages} {1} (\bibinfo {year} {1962})}\BibitemShut
  {NoStop}%
\bibitem [{\citenamefont {Pippard}(1964)}]{pippard1964}%
  \BibitemOpen
  \bibfield  {author} {\bibinfo {author} {\bibfnamefont {A.~B.}\ \bibnamefont
  {Pippard}},\ }\bibfield  {title} {\bibinfo {title} {Quantization of {{Coupled
  Orbits}} in {{Metals II}}. {{The Two-Dimensional Network}}, with {{Special
  Reference}} to the {{Properties}} of {{Zinc}}},\ }\href
  {https://doi.org/https://doi.org/10.1098/rsta.1964.0008} {\bibfield
  {journal} {\bibinfo  {journal} {Philosophical Transactions of the Royal
  Society of London. Series A, Mathematical and Physical Sciences}\ }\textbf
  {\bibinfo {volume} {256}},\ \bibinfo {pages} {317} (\bibinfo {year}
  {1964})}\BibitemShut {NoStop}%
\bibitem [{\citenamefont {Chambers}(1966)}]{chambersMB1966}%
  \BibitemOpen
  \bibfield  {author} {\bibinfo {author} {\bibfnamefont {R.~G.}\ \bibnamefont
  {Chambers}},\ }\bibfield  {title} {\bibinfo {title} {Magnetic breakdown in
  real metals},\ }\href {https://doi.org/10.1088/0370-1328/88/3/318} {\bibfield
   {journal} {\bibinfo  {journal} {Proceedings of the Physical Society}\
  }\textbf {\bibinfo {volume} {88}},\ \bibinfo {pages} {701} (\bibinfo {year}
  {1966})}\BibitemShut {NoStop}%
\bibitem [{\citenamefont {Alexandradinata}\ and\ \citenamefont
  {Glazman}(2017)}]{alexandradinataMB2017}%
  \BibitemOpen
  \bibfield  {author} {\bibinfo {author} {\bibfnamefont {A.}~\bibnamefont
  {Alexandradinata}}\ and\ \bibinfo {author} {\bibfnamefont {L.}~\bibnamefont
  {Glazman}},\ }\bibfield  {title} {\bibinfo {title} {Geometric {{Phase}} and
  {{Orbital Moment}} in {{Quantization Rules}} for {{Magnetic Breakdown}}},\
  }\href {https://doi.org/10.1103/PhysRevLett.119.256601} {\bibfield  {journal}
  {\bibinfo  {journal} {Physical Review Letters}\ }\textbf {\bibinfo {volume}
  {119}},\ \bibinfo {pages} {256601} (\bibinfo {year} {2017})}\BibitemShut
  {NoStop}%
\bibitem [{\citenamefont {Alexandradinata}\ and\ \citenamefont
  {Glazman}(2018)}]{alexandradinataModernMB2018}%
  \BibitemOpen
  \bibfield  {author} {\bibinfo {author} {\bibfnamefont {A.}~\bibnamefont
  {Alexandradinata}}\ and\ \bibinfo {author} {\bibfnamefont {L.}~\bibnamefont
  {Glazman}},\ }\bibfield  {title} {\bibinfo {title} {Modern semiclassical
  theory of magnetic transport and breakdown},\ }\href@noop {} {\bibfield
  {journal} {\bibinfo  {journal} {Physical Review B}\ }\textbf {\bibinfo
  {volume} {97}},\ \bibinfo {pages} {144422} (\bibinfo {year} {2018})},\
  \Eprint {https://arxiv.org/abs/1710.04215} {arXiv:1710.04215} \BibitemShut
  {NoStop}%
\bibitem [{\citenamefont {Stark}\ and\ \citenamefont
  {Friedberg}(1971)}]{starkMB1971a}%
  \BibitemOpen
  \bibfield  {author} {\bibinfo {author} {\bibfnamefont {R.~W.}\ \bibnamefont
  {Stark}}\ and\ \bibinfo {author} {\bibfnamefont {C.~B.}\ \bibnamefont
  {Friedberg}},\ }\bibfield  {title} {\bibinfo {title} {Quantum
  {{Interference}} of {{Electron Waves}} in a {{Normal Metal}}},\ }\href
  {https://doi.org/10.1103/PhysRevLett.26.556} {\bibfield  {journal} {\bibinfo
  {journal} {Physical Review Letters}\ }\textbf {\bibinfo {volume} {26}},\
  \bibinfo {pages} {556} (\bibinfo {year} {1971})}\BibitemShut {NoStop}%
\bibitem [{\citenamefont {Lu}\ and\ \citenamefont
  {Fertig}(2014)}]{luMBTBG2014}%
  \BibitemOpen
  \bibfield  {author} {\bibinfo {author} {\bibfnamefont {C.-K.}\ \bibnamefont
  {Lu}}\ and\ \bibinfo {author} {\bibfnamefont {H.~A.}\ \bibnamefont
  {Fertig}},\ }\bibfield  {title} {\bibinfo {title} {Magnetic {{Breakdown}} in
  {{Twisted Bilayer Graphene}}},\ }\href
  {https://doi.org/10.1103/PhysRevB.89.085408} {\bibfield  {journal} {\bibinfo
  {journal} {Physical Review B}\ }\textbf {\bibinfo {volume} {89}},\ \bibinfo
  {pages} {085408} (\bibinfo {year} {2014})}\BibitemShut {NoStop}%
\bibitem [{\citenamefont {Hejazi}\ \emph {et~al.}(2019)\citenamefont {Hejazi},
  \citenamefont {Liu},\ and\ \citenamefont {Balents}}]{hejaziTwisted2019}%
  \BibitemOpen
  \bibfield  {author} {\bibinfo {author} {\bibfnamefont {K.}~\bibnamefont
  {Hejazi}}, \bibinfo {author} {\bibfnamefont {C.}~\bibnamefont {Liu}},\ and\
  \bibinfo {author} {\bibfnamefont {L.}~\bibnamefont {Balents}},\ }\bibfield
  {title} {\bibinfo {title} {Landau levels in twisted bilayer graphene and
  semiclassical orbits},\ }\href {https://doi.org/10.1103/PhysRevB.100.035115}
  {\bibfield  {journal} {\bibinfo  {journal} {Physical Review B}\ }\textbf
  {\bibinfo {volume} {100}},\ \bibinfo {pages} {035115} (\bibinfo {year}
  {2019})}\BibitemShut {NoStop}%
\bibitem [{\citenamefont {Paul}\ \emph {et~al.}(2022)\citenamefont {Paul},
  \citenamefont {Crowley}, \citenamefont {Devakul},\ and\ \citenamefont
  {Fu}}]{paulMoireLandauFans2022}%
  \BibitemOpen
  \bibfield  {author} {\bibinfo {author} {\bibfnamefont {N.}~\bibnamefont
  {Paul}}, \bibinfo {author} {\bibfnamefont {P.~J.~D.}\ \bibnamefont
  {Crowley}}, \bibinfo {author} {\bibfnamefont {T.}~\bibnamefont {Devakul}},\
  and\ \bibinfo {author} {\bibfnamefont {L.}~\bibnamefont {Fu}},\ }\bibfield
  {title} {\bibinfo {title} {Moir\'e {{Landau Fans}} and {{Magic Zeros}}},\
  }\href {https://doi.org/10.1103/PhysRevLett.129.116804} {\bibfield  {journal}
  {\bibinfo  {journal} {Physical Review Letters}\ }\textbf {\bibinfo {volume}
  {129}},\ \bibinfo {pages} {116804} (\bibinfo {year} {2022})}\BibitemShut
  {NoStop}%
\bibitem [{\citenamefont {Paul}\ \emph {et~al.}(2024)\citenamefont {Paul},
  \citenamefont {Crowley},\ and\ \citenamefont {Fu}}]{paulMB2024}%
  \BibitemOpen
  \bibfield  {author} {\bibinfo {author} {\bibfnamefont {N.}~\bibnamefont
  {Paul}}, \bibinfo {author} {\bibfnamefont {P.~J.~D.}\ \bibnamefont
  {Crowley}},\ and\ \bibinfo {author} {\bibfnamefont {L.}~\bibnamefont {Fu}},\
  }\bibfield  {title} {\bibinfo {title} {Directional {{Localization}} from a
  {{Magnetic Field}} in {{Moir\'e Systems}}},\ }\href
  {https://doi.org/10.1103/PhysRevLett.132.246402} {\bibfield  {journal}
  {\bibinfo  {journal} {Physical Review Letters}\ }\textbf {\bibinfo {volume}
  {132}},\ \bibinfo {pages} {246402} (\bibinfo {year} {2024})}\BibitemShut
  {NoStop}%
\bibitem [{\citenamefont {Bocarsly}\ \emph {et~al.}(2024)\citenamefont
  {Bocarsly}, \citenamefont {Uzan}, \citenamefont {Roy}, \citenamefont
  {Grover}, \citenamefont {Xiao}, \citenamefont {Dong}, \citenamefont
  {Labendik}, \citenamefont {Uri}, \citenamefont {Huber}, \citenamefont
  {Myasoedov}, \citenamefont {Watanabe}, \citenamefont {Taniguchi},
  \citenamefont {Yan}, \citenamefont {Levitov},\ and\ \citenamefont
  {Zeldov}}]{bocarslyHaasVanAlphen2024}%
  \BibitemOpen
  \bibfield  {author} {\bibinfo {author} {\bibfnamefont {M.}~\bibnamefont
  {Bocarsly}}, \bibinfo {author} {\bibfnamefont {M.}~\bibnamefont {Uzan}},
  \bibinfo {author} {\bibfnamefont {I.}~\bibnamefont {Roy}}, \bibinfo {author}
  {\bibfnamefont {S.}~\bibnamefont {Grover}}, \bibinfo {author} {\bibfnamefont
  {J.}~\bibnamefont {Xiao}}, \bibinfo {author} {\bibfnamefont {Z.}~\bibnamefont
  {Dong}}, \bibinfo {author} {\bibfnamefont {M.}~\bibnamefont {Labendik}},
  \bibinfo {author} {\bibfnamefont {A.}~\bibnamefont {Uri}}, \bibinfo {author}
  {\bibfnamefont {M.~E.}\ \bibnamefont {Huber}}, \bibinfo {author}
  {\bibfnamefont {Y.}~\bibnamefont {Myasoedov}}, \bibinfo {author}
  {\bibfnamefont {K.}~\bibnamefont {Watanabe}}, \bibinfo {author}
  {\bibfnamefont {T.}~\bibnamefont {Taniguchi}}, \bibinfo {author}
  {\bibfnamefont {B.}~\bibnamefont {Yan}}, \bibinfo {author} {\bibfnamefont
  {L.~S.}\ \bibnamefont {Levitov}},\ and\ \bibinfo {author} {\bibfnamefont
  {E.}~\bibnamefont {Zeldov}},\ }\bibfield  {title} {\bibinfo {title} {De
  {{Haas}}--van {{Alphen}} spectroscopy and magnetic breakdown in moir\'e
  graphene},\ }\href {https://doi.org/10.1126/science.adh3499} {\bibfield
  {journal} {\bibinfo  {journal} {Science}\ }\textbf {\bibinfo {volume}
  {383}},\ \bibinfo {pages} {42} (\bibinfo {year} {2024})}\BibitemShut
  {NoStop}%
\bibitem [{\citenamefont {Auerbach}\ \emph {et~al.}(2025)\citenamefont
  {Auerbach}, \citenamefont {Dutta}, \citenamefont {Uzan}, \citenamefont
  {Vituri}, \citenamefont {Zhou}, \citenamefont {Meltzer}, \citenamefont
  {Grover}, \citenamefont {Holder}, \citenamefont {Emanuel}, \citenamefont
  {Huber}, \citenamefont {Myasoedov}, \citenamefont {Watanabe}, \citenamefont
  {Taniguchi}, \citenamefont {Oreg}, \citenamefont {Berg},\ and\ \citenamefont
  {Zeldov}}]{auerbachIsospin2025}%
  \BibitemOpen
  \bibfield  {author} {\bibinfo {author} {\bibfnamefont {N.}~\bibnamefont
  {Auerbach}}, \bibinfo {author} {\bibfnamefont {S.}~\bibnamefont {Dutta}},
  \bibinfo {author} {\bibfnamefont {M.}~\bibnamefont {Uzan}}, \bibinfo {author}
  {\bibfnamefont {Y.}~\bibnamefont {Vituri}}, \bibinfo {author} {\bibfnamefont
  {Y.}~\bibnamefont {Zhou}}, \bibinfo {author} {\bibfnamefont {A.~Y.}\
  \bibnamefont {Meltzer}}, \bibinfo {author} {\bibfnamefont {S.}~\bibnamefont
  {Grover}}, \bibinfo {author} {\bibfnamefont {T.}~\bibnamefont {Holder}},
  \bibinfo {author} {\bibfnamefont {P.}~\bibnamefont {Emanuel}}, \bibinfo
  {author} {\bibfnamefont {M.~E.}\ \bibnamefont {Huber}}, \bibinfo {author}
  {\bibfnamefont {Y.}~\bibnamefont {Myasoedov}}, \bibinfo {author}
  {\bibfnamefont {K.}~\bibnamefont {Watanabe}}, \bibinfo {author}
  {\bibfnamefont {T.}~\bibnamefont {Taniguchi}}, \bibinfo {author}
  {\bibfnamefont {Y.}~\bibnamefont {Oreg}}, \bibinfo {author} {\bibfnamefont
  {E.}~\bibnamefont {Berg}},\ and\ \bibinfo {author} {\bibfnamefont
  {E.}~\bibnamefont {Zeldov}},\ }\bibfield  {title} {\bibinfo {title} {Isospin
  magnetic texture and intervalley exchange interaction in rhombohedral
  tetralayer graphene},\ }\href {https://doi.org/10.1038/s41567-025-03035-z}
  {\bibfield  {journal} {\bibinfo  {journal} {Nature Physics}\ }\textbf
  {\bibinfo {volume} {21}},\ \bibinfo {pages} {1765} (\bibinfo {year}
  {2025})}\BibitemShut {NoStop}%
\bibitem [{\citenamefont {Ando}\ and\ \citenamefont {Uemura}(1974)}]{ando1974}%
  \BibitemOpen
  \bibfield  {author} {\bibinfo {author} {\bibfnamefont {T.}~\bibnamefont
  {Ando}}\ and\ \bibinfo {author} {\bibfnamefont {Y.}~\bibnamefont {Uemura}},\
  }\bibfield  {title} {\bibinfo {title} {Theory of {{Quantum Transport}} in a
  {{Two-Dimensional Electron System}} under {{Magnetic Fields}}. {{I}}.
  {{Characteristics}} of {{Level Broadening}} and {{Transport}} under {{Strong
  Fields}}},\ }\href {https://doi.org/10.1143/JPSJ.36.959} {\bibfield
  {journal} {\bibinfo  {journal} {Journal of the Physical Society of Japan}\
  }\textbf {\bibinfo {volume} {36}},\ \bibinfo {pages} {959} (\bibinfo {year}
  {1974})}\BibitemShut {NoStop}%
\bibitem [{\citenamefont {Ando}(1974)}]{ando1974a}%
  \BibitemOpen
  \bibfield  {author} {\bibinfo {author} {\bibfnamefont {T.}~\bibnamefont
  {Ando}},\ }\bibfield  {title} {\bibinfo {title} {Theory of {{Quantum
  Transport}} in a {{Two-Dimensional Electron System}} under {{Magnetic
  Fields}}. {{IV}}. {{Oscillatory Conductivity}}},\ }\href
  {https://doi.org/10.1143/JPSJ.37.1233} {\bibfield  {journal} {\bibinfo
  {journal} {Journal of the Physical Society of Japan}\ }\textbf {\bibinfo
  {volume} {37}},\ \bibinfo {pages} {1233} (\bibinfo {year}
  {1974})}\BibitemShut {NoStop}%
\bibitem [{\citenamefont {Ando}\ \emph {et~al.}(1982)\citenamefont {Ando},
  \citenamefont {Fowler},\ and\ \citenamefont {Stern}}]{andoRMP1982a}%
  \BibitemOpen
  \bibfield  {author} {\bibinfo {author} {\bibfnamefont {T.}~\bibnamefont
  {Ando}}, \bibinfo {author} {\bibfnamefont {A.~B.}\ \bibnamefont {Fowler}},\
  and\ \bibinfo {author} {\bibfnamefont {F.}~\bibnamefont {Stern}},\ }\bibfield
   {title} {\bibinfo {title} {Electronic properties of two-dimensional
  systems},\ }\href {https://doi.org/10.1103/RevModPhys.54.437} {\bibfield
  {journal} {\bibinfo  {journal} {Reviews of Modern Physics}\ }\textbf
  {\bibinfo {volume} {54}},\ \bibinfo {pages} {437} (\bibinfo {year}
  {1982})}\BibitemShut {NoStop}%
\bibitem [{SM()}]{SM}%
  \BibitemOpen
  \href@noop {} {}\bibinfo {note} {See supplemental material for derivations
  and extended discussions.}\BibitemShut {Stop}%
\bibitem [{\citenamefont {Min}\ and\ \citenamefont
  {MacDonald}(2008)}]{minABC2008}%
  \BibitemOpen
  \bibfield  {author} {\bibinfo {author} {\bibfnamefont {H.}~\bibnamefont
  {Min}}\ and\ \bibinfo {author} {\bibfnamefont {A.~H.}\ \bibnamefont
  {MacDonald}},\ }\bibfield  {title} {\bibinfo {title} {Chiral decomposition in
  the electronic structure of graphene multilayers},\ }\href
  {https://doi.org/10.1103/PhysRevB.77.155416} {\bibfield  {journal} {\bibinfo
  {journal} {Physical Review B}\ }\textbf {\bibinfo {volume} {77}},\ \bibinfo
  {pages} {155416} (\bibinfo {year} {2008})}\BibitemShut {NoStop}%
\bibitem [{\citenamefont {Koshino}\ and\ \citenamefont
  {McCann}(2009)}]{koshinoTrigonal2009}%
  \BibitemOpen
  \bibfield  {author} {\bibinfo {author} {\bibfnamefont {M.}~\bibnamefont
  {Koshino}}\ and\ \bibinfo {author} {\bibfnamefont {E.}~\bibnamefont
  {McCann}},\ }\bibfield  {title} {\bibinfo {title} {Trigonal warping and
  {{Berry}}'s phase {{N}} {$\pi$} in {{ABC-stacked}} multilayer graphene},\
  }\href {https://doi.org/10.1103/PhysRevB.80.165409} {\bibfield  {journal}
  {\bibinfo  {journal} {Physical Review B}\ }\textbf {\bibinfo {volume} {80}},\
  \bibinfo {pages} {165409} (\bibinfo {year} {2009})}\BibitemShut {NoStop}%
\bibitem [{\citenamefont {Shoenberg}(1984)}]{shoenbergSdH1984a}%
  \BibitemOpen
  \bibfield  {author} {\bibinfo {author} {\bibfnamefont {D.}~\bibnamefont
  {Shoenberg}},\ }\href@noop {} {\emph {\bibinfo {title} {Magnetic
  {{Oscillations}} in {{Metals}}}}}\ (\bibinfo  {publisher} {Cambridge
  University Press},\ \bibinfo {year} {1984})\BibitemShut {NoStop}%
\bibitem [{\citenamefont {Mikitik}\ and\ \citenamefont
  {Sharlai}(1999)}]{mikitikBerrysPhase1999}%
  \BibitemOpen
  \bibfield  {author} {\bibinfo {author} {\bibfnamefont {G.~P.}\ \bibnamefont
  {Mikitik}}\ and\ \bibinfo {author} {\bibfnamefont {{\relax Yu}.~V.}\
  \bibnamefont {Sharlai}},\ }\bibfield  {title} {\bibinfo {title}
  {Manifestation of {{Berry}}'s {{Phase}} in {{Metal Physics}}},\ }\href
  {https://doi.org/10.1103/PhysRevLett.82.2147} {\bibfield  {journal} {\bibinfo
   {journal} {Physical Review Letters}\ }\textbf {\bibinfo {volume} {82}},\
  \bibinfo {pages} {2147} (\bibinfo {year} {1999})}\BibitemShut {NoStop}%
\bibitem [{Note1()}]{Note1}%
  \BibitemOpen
  \bibinfo {note} {Strictly speaking, Shubnikov--de Haas oscillations refer to
  quantum oscillations in longitudinal transport. Throughout this work, we
  therefore use ``SdH oscillations'' only for the oscillations of $\rho _{xx}$,
  while the corresponding oscillations in the density of states are referred to
  as ``DOS oscillations''.}\BibitemShut {Stop}%
\bibitem [{Ju_()}]{Ju_private}%
  \BibitemOpen
  \href@noop {} {}\bibinfo {note} {L. Ju, private communication
  (2026).}\BibitemShut {Stop}%
\bibitem [{Note2()}]{Note2}%
  \BibitemOpen
  \bibinfo {note} {For a chiral $p+ip$ state, the BdG Chern-number transition
  is controlled by whether the chemical potential lies above or below
  $E(\protect \mathbf {k}=0)$ \cite
  {chouQM2026,ghoshNonAbelian2010}.}\BibitemShut {Stop}%
\bibitem [{\citenamefont {Ghosh}\ \emph {et~al.}(2010)\citenamefont {Ghosh},
  \citenamefont {Sau}, \citenamefont {Tewari},\ and\ \citenamefont
  {Das~Sarma}}]{ghoshNonAbelian2010}%
  \BibitemOpen
  \bibfield  {author} {\bibinfo {author} {\bibfnamefont {P.}~\bibnamefont
  {Ghosh}}, \bibinfo {author} {\bibfnamefont {J.~D.}\ \bibnamefont {Sau}},
  \bibinfo {author} {\bibfnamefont {S.}~\bibnamefont {Tewari}},\ and\ \bibinfo
  {author} {\bibfnamefont {S.}~\bibnamefont {Das~Sarma}},\ }\bibfield  {title}
  {\bibinfo {title} {Non-{{Abelian}} topological order in noncentrosymmetric
  superconductors with broken time-reversal symmetry},\ }\href
  {https://doi.org/10.1103/PhysRevB.82.184525} {\bibfield  {journal} {\bibinfo
  {journal} {Physical Review B}\ }\textbf {\bibinfo {volume} {82}},\ \bibinfo
  {pages} {184525} (\bibinfo {year} {2010})}\BibitemShut {NoStop}%
\bibitem [{\citenamefont {Huang}\ and\ \citenamefont
  {Das~Sarma}(2026)}]{huangDisorder2026}%
  \BibitemOpen
  \bibfield  {author} {\bibinfo {author} {\bibfnamefont {Y.}~\bibnamefont
  {Huang}}\ and\ \bibinfo {author} {\bibfnamefont {S.}~\bibnamefont
  {Das~Sarma}},\ }\bibfield  {title} {\bibinfo {title} {Disorder-induced
  strong-field strong localization in two-dimensional systems},\ }\href
  {https://doi.org/10.1103/7dq6-dh9z} {\bibfield  {journal} {\bibinfo
  {journal} {Physical Review B}\ }\textbf {\bibinfo {volume} {113}},\ \bibinfo
  {pages} {184203} (\bibinfo {year} {2026})}\BibitemShut {NoStop}%
\bibitem [{\citenamefont {Dingle}(1952)}]{dingle1952}%
  \BibitemOpen
  \bibfield  {author} {\bibinfo {author} {\bibfnamefont {R.~B.}\ \bibnamefont
  {Dingle}},\ }\bibfield  {title} {\bibinfo {title} {Some magnetic properties
  of metals {{II}}. {{The}} influence of collisions on the magnetic behaviour
  of large systems},\ }\href {https://doi.org/10.1098/rspa.1952.0056}
  {\bibfield  {journal} {\bibinfo  {journal} {Proceedings of the Royal Society
  of London. A. Mathematical and Physical Sciences}\ }\textbf {\bibinfo
  {volume} {211}},\ \bibinfo {pages} {517} (\bibinfo {year}
  {1952})}\BibitemShut {NoStop}%
\end{thebibliography}%

\bigskip
\bigskip
\begin{center}
\textbf{\large End Matter}
\end{center}
\bigskip
\vspace{-0.2cm}

Here we provide details of the transport calculation we used to obtain the longitudinal resistivity. 
We calculate the disorder-average Green's function with the self-consistent Born approximation (SCBA) \cite{ando1974, ando1974a, andoRMP1982a, huangDisorder2026} (see SM for a discussion \cite{SM})
\begin{equation} \label{eqn:SCBA_equations}
    \Sigma_N(E,B) = \frac{\Gamma(B)^2}{4}\sum_{N'}G_{N'}(E,B), 
\end{equation}
\begin{equation} \label{eqn:SCBA_equations_Gn}
    G_N(E,B) = \frac{1}{E-E_N-\Sigma_N(E,B)}~, 
\end{equation}
where the subscript $N$ denotes the Landau level index. 
The broadening is characterized by $\Gamma^2(B)/4=n_{\mathrm{imp}}u_0^2/(2\pi l_B^2)$, where $n_{\mathrm{imp}}$ and $u_0$ are the impurity density and disorder strength, respectively. The factor $1/(2\pi l_B^2)\propto B$ comes from the Landau-level degeneracy per unit area so that $\Gamma(B)\propto\sqrt{B}$. 
We parametrize the overall disorder strength by an effective Dingle temperature $T_D$, defined through $\pi k_B T_D=\Gamma(B=1~\mathrm{T})$.
For a typical $T_D=0.05$K used in our calculation, a rough estimation through $\tau_q=\hbar/(2\Gamma)$ gives $\tau_q\approx 24$ ps, a bit cleaner than the experimental fit Ref.\cite{kalantreFermiology2026} with $\tau_q\approx 4$ ps. 
We note that the R4G systems are ultra-clean, and the inclusion of disorder is mainly for broadening the delta functions in the clean limit. Using a Lorentzian broadening (e.g., direct application of Dingle temperature) gives qualitatively similar results as discussed in SM \cite{SM}.

\begin{figure}[t]
    \centering
    \includegraphics[width=0.99\linewidth]{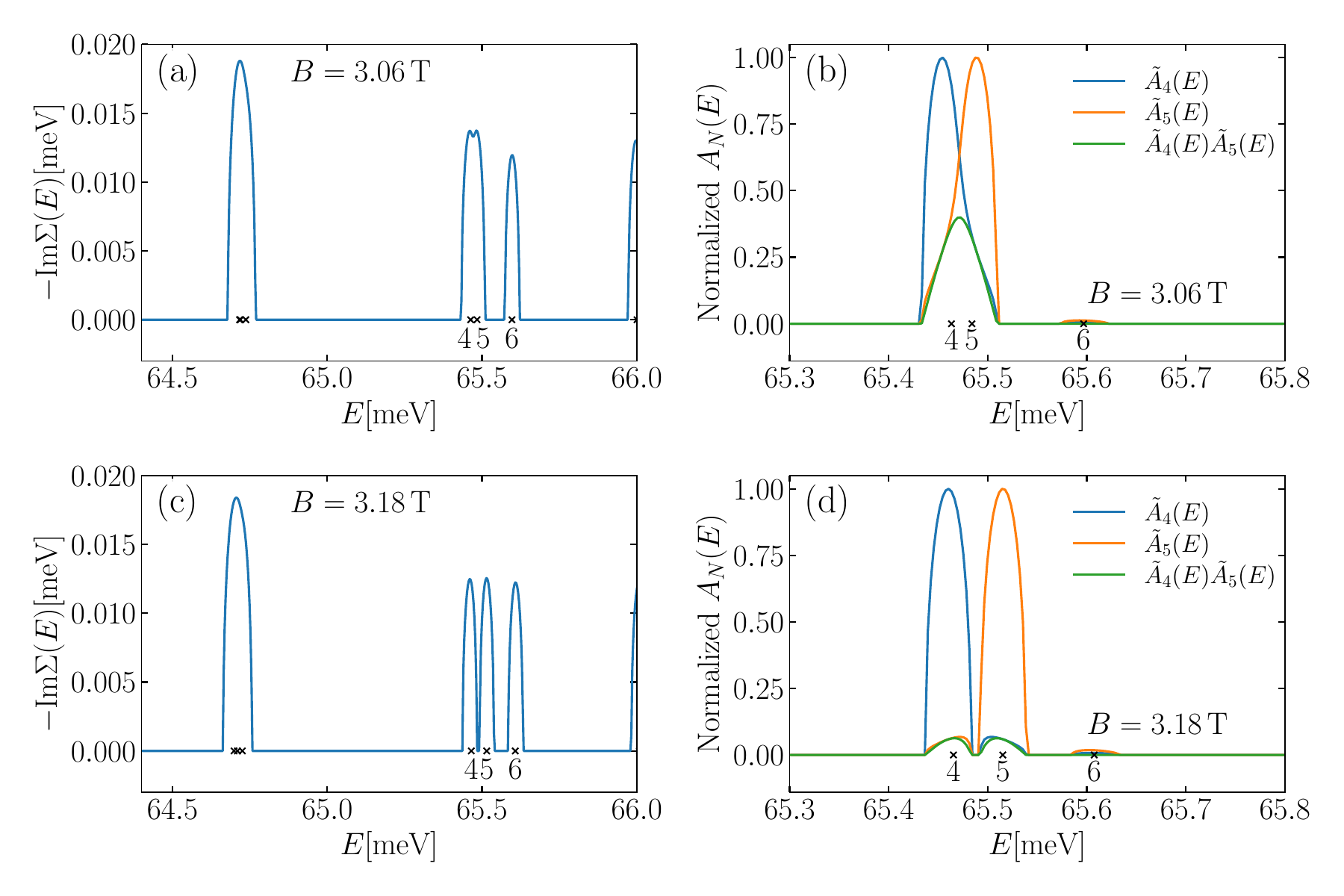}
    \caption{Examples of convergent self-energy and the corresponding spectrum function at different magnetic fields, for displacement field $u_D=43$ meV and disorder broadening $T_D=0.05$ K. 
    Note the self-energy in the SCBA scheme Eq.~\eqref{eqn:SCBA_equations} is independent of $N$. 
    (a) $-\mathrm{Im}\Sigma(E)$ near the six lowest Landau level energies at $B=3.07$ T. 
    The corresponding Landau level energies are marked by black crosses. 
    (b) Normalized spectrum function $\tilde{A}_4(E)$ and $\tilde{A}_5(E)$ for the fourth and fifth Landau levels,  together with their product $\tilde{A}_4(E)\tilde{A}_5(E)$.   
    Each spectral function is normalized to unit maximum. 
    The product $\tilde{A}_4(E)\tilde{A}_5(E)$ reaches its maximum between the two Landau-level energies as a result of a substantial spectral overlap.
    (c) Same as panel (a), but at $B=3.47$ T.
    (d) Same as panel (b), but at $B=3.47$ T. 
    In this case, the fourth and fifth Landau levels are separated, and the overlap between $\tilde{A}_4(E)$ and $\tilde{A}_5(E)$ is substantially reduced.
    }
    \label{fig:SCBA}
\end{figure}

Using the SCBA Green's function, we calculate the thermally broadened DOS $D(\mu, B)$ and the longitudinal conductivity $\sigma_{xx}(\mu, B)$ \cite{ando1974, ando1974a, andoRMP1982a} within the bubble approximation: 
\begin{equation}
    D(\mu,B)=\frac{1}{2\pi l_B^2}\int \mathrm{d}E\frac{-\partial f(E-\mu)}{\partial E}\sum_N A_N(E,B)~,
\end{equation}
\begin{equation}\label{eqn:conductivity}
\begin{aligned}
    \sigma_{xx}(\mu,B) = &\frac{e^2}{2\hbar l_B^2} \int \mathrm{d}E \frac{-\partial f(E-\mu)}{\partial E} \\
    &\times \sum_{NN'}|v^x_{NN'}|^2 A_{N}(E,B) A_{N'}(E,B)~,
\end{aligned}
\end{equation}
where $f(E)$ is the Fermi distribution function, $v^x_{N'N} = \langle N'| \partial H/\partial k_x |N\rangle$  is the velocity vertex, 
$A_N(E,B)=-\frac{1}{\pi}\mathrm{Im}G_N(E,B)$ is the spectral functions of the $N$-th Landau level.  
We neglect vertex corrections to the conductivity and retain only the bubble contribution.  While such corrections may quantitatively modify $\sigma_{xx}$, the present approximation is sufficient to capture the magnetic breakdown-induced oscillation patterns discussed in the main text. 
The longitudinal $\rho_{xx}$ is obtained as $\rho_{xx} \sim \sigma_{xx}/(\sigma_{xx}^2+\sigma_{xy}^2)$, with $\sigma_{xy}$ approximated as its semiclassical value $\sigma_{xy} = ne/B$. 
Following experiments, we transform the resistivity into a function of carrier density $n$, using $n(\mu,B)=\int_0^\mu \mathrm{d}\mu'D(\mu',B)$ measured relative to charge neutrality.

The conductivity in Eq.~\eqref{eqn:conductivity} is controlled by products of spectral functions from pairs of Landau levels weighted by the corresponding velocity matrix elements. 
For dispersionless Landau levels, the diagonal matrix elements ${v}^{x}_{NN}$ vanish, so the dominant contributions arise from $N\neq N'$. 
Consequently, the longitudinal conductivity is strongly enhanced when two Landau levels with a nonzero velocity matrix element approach one another and their disorder-broadened spectral functions overlap, as indicated in Fig.~\ref{fig:SCBA}. 
This explains why the Landau-level crossing and reconnection structure leaves a more pronounced signature in the SdH oscillations of the resistivity than in the DOS oscillations.


\newpage \clearpage 
	
	\onecolumngrid
	
	\begin{center}
		{\large
			Magnetic Breakdown Induced Anomalous Quantum Oscillation in Rhombohedral Tetralayer Graphene
			\vspace{4pt}
			\\
			SUPPLEMENTAL MATERIAL
		}
	\end{center}
	
	\setcounter{figure}{0}
	\renewcommand{\thefigure}{S\arabic{figure}}
	\setcounter{equation}{0}
	\renewcommand{\theequation}{S\arabic{equation}}

	In this supplemental material, we provide technical details and additional results supporting the main results presented in the main text.

\section{Continuum model and numerical parameters}

We take the noninteracting continuum Hamiltonian of rhombohedral tetralayer graphene as 
\begin{equation}
    H_0 = \sum_{\eta=\pm1}\sum_{\mathbf{k}}
    \Psi_{\eta,\mathbf{k}}^{\dagger}
    h_{\eta}(\mathbf{k})
    \Psi_{\eta,\mathbf{k}},
\end{equation}
where $\eta=+1$ and $-1$ label valleys $K$ and $K'$, respectively, and $\mathbf{k}$ is measured relative to the corresponding valley center. 
$\Psi_{\eta;\mathbf{k}}$ is an eight component spinor written in the basis $(A_1,B_1,A_2,B_2,A_3,B_3,A_4,B_4)$. 
Here the single-particle Hamiltonian is 
\begin{equation} \label{eqn:R4G_hamiltonian}
 h_{\eta}(\mathbf{k})=
\begin{pmatrix}
-\frac{3}{2}u_D & v_0\Pi_\eta^\dagger & v_4\Pi_\eta^\dagger & v_3\Pi_\eta & 0 & \frac{\gamma_2}{2} & 0 & 0 \\
v_0\Pi_\eta & -\frac{3}{2}u_D & \gamma_1 & v_4\Pi_\eta^\dagger & 0 & 0 & 0 & 0 \\
v_4\Pi_\eta & \gamma_1 & -\frac{1}{2}u_D & v_0\Pi_\eta^\dagger & v_4\Pi_\eta^\dagger & v_3\Pi_\eta & 0 & \frac{\gamma_2}{2} \\
v_3\Pi_\eta^\dagger & v_4\Pi_\eta & v_0\Pi_\eta & -\frac{1}{2}u_D & \gamma_1 & v_4\Pi_\eta^\dagger & 0 & 0 \\
0 & 0 & v_4\Pi_\eta & \gamma_1 & \frac{1}{2}u_D & v_0\Pi_\eta^\dagger & v_4\Pi_\eta^\dagger & v_3\Pi_\eta \\
\frac{\gamma_2}{2} & 0 & v_3\Pi_\eta^\dagger & v_4\Pi_\eta & v_0\Pi_\eta & \frac{1}{2}u_D & \gamma_1 & v_4\Pi_\eta^\dagger \\
0 & 0 & 0 & 0 & v_4\Pi_\eta & \gamma_1 & \frac{3}{2}u_D & v_0\Pi_\eta^\dagger \\
0 & 0 & \frac{\gamma_2}{2} & 0 & v_3\Pi_\eta^\dagger & v_4\Pi_\eta & v_0\Pi_\eta & \frac{3}{2}u_D
\end{pmatrix}+ \mathcal{U}_0~,
\end{equation}
\begin{align}
    \mathcal{U}_0 &= \mathrm{diag} \left[ \Delta_2, \Delta_2+\delta, -\Delta_2+\delta, -\Delta_2+\delta, -\Delta_2+\delta, -\Delta_2+\delta, \Delta_2+\delta, \Delta_2 \right]~,
\end{align}
where $\Pi_\eta=\eta k_x+ik_y$ and $ v_j=\frac{\sqrt{3}}{2}\gamma_j a$, $a= 2.46\,\text{\AA}$ is lattice constant of graphene.
In our calculation, we focus on the $K$ valley and follow Ref.~\cite{kalantreFermiology2026, auerbachIsospin2025} to take the parameters as $\gamma_0 = 3100 $ meV, $\gamma_1=380$ meV,  $\gamma_2 = -15$ meV,  $\gamma_3=-290$ meV, $\gamma_4 =-141$ meV, 
$\delta = 10.5 $ meV and 
$\Delta_2 = 2$ meV. 

To derive the Landau levels, we introduce the perpendicular magnetic field through the Peierls substitution
$k_i\rightarrow \hat\pi_{i} \equiv -i\partial_i+eA_i$, 
which satisfy $[\hat\pi_x,\hat\pi_y]=-i/l_B^2$. 
We then introduce the ladder operators 
\begin{equation}
    \hat{a} = i\frac{l_B}{\sqrt{2}}(\hat \pi_x - i\hat\pi_y),\qquad
    \hat{a}^\dagger = -i\frac{l_B}{\sqrt{2}}(\hat \pi_x + i\hat\pi_y). 
\end{equation}
which satisfy $[\hat{a},\hat{a}^\dagger]=1$. 
The Hamiltonian can then be solved exactly within a finite cutoff of Landau levels.

We expand each layer and sublattice component in the harmonic-oscillator basis and retain the states $|n\rangle$ with $n=0,\ldots, N_l^\alpha-1$. 
Here we take the cutoff to be $N_{l}^A = N_L+l$ for the $A$ site and $N_l^B = N_L+l+1$ for the $B$ site, where $l=1,2,3,4$ denotes the layer index. 
The offsets are introduced to avoid spurious states at the truncation boundary. 
In all calculations presented in the main text, we take $N_L=250$.

For later use, we also define the velocity 
\begin{equation}\label{eqn:velocity}
    {v}_x(\mathbf{k}) = \frac{\partial h_\eta(\mathbf{k})}{\partial k_x},\qquad
    {v}_y(\mathbf{k}) = \frac{\partial h_\eta(\mathbf{k})}{\partial k_y}~.
\end{equation}
Under the same Peierls substitution, the velocities become operators within the harmonic-oscillator basis
\begin{equation}
    v_x(\mathbf{k})\rightarrow \hat{v}_x(\hat{\boldsymbol{\pi}}), \qquad
    v_y(\mathbf{k})\rightarrow \hat{v}_y(\hat{\boldsymbol{\pi}})~.
\end{equation}

\section{Calculation of longitudinal resistivity}

\subsection{Longitudinal conductivity}
\label{app:conductivity}

For completeness, we derive the Kubo formula of longitudinal conductivity. 
We define the velocity-vertex matrix elements in the Landau-level basis as
\begin{equation} \label{eqn:current_operator}
    v^x_{N'N} \equiv \left\langle N' \left|
    \hat{v}_x(\hat{\boldsymbol{\pi}})
    \right| N \right\rangle ,\quad 
    {J}_x = -\frac{e}{\hbar\mathcal{A}} \sum_{X,N',N} v^x_{N'N} c^\dagger_{N'X}c_{NX},
\end{equation}
where $N,N'$ label Landau levels, $X$ labels the guiding-center degeneracy $\mathcal{A}$ is the sample area, 
and $J_x$ is the current density.
In numerical calculations, velocity elements are retained within an energy window of $|E_{N'}-E_N|<10~\mathrm{meV}$.

The longitudinal conductivity can be obtained from the Kubo formula
\begin{equation} \label{eqn:conductivity_from_Pi}
    \sigma_{xx}(\omega) =
    \frac{K_{xx}(\omega)-K_{xx}(0)}{i\omega},
\end{equation}
where $K_{xx}$ is the retarded current--current correlation function. 
Within the bubble approximation, the corresponding Matsubara correlation function is
\begin{equation} \label{eqn:Kxx}
    K_{xx}(i\omega_{\ell},B) =
    -\frac{e^2}{\beta\hbar^2\mathcal{A}}
    \sum_{\varepsilon_j}
    \sum_{N, N', X}
    v^x_{N'N}v^x_{NN'}
    G_N(i\varepsilon_j+i\omega_{\ell},B)
    G_{N'}(i\varepsilon_j,B)~,
\end{equation}
where $i\varepsilon_j$ and $i\omega_{\ell}$ are fermionic and bosonic Matsubara frequencies, respectively. 
$G_N(i\varepsilon_j)$ is the Green's function of electron on the $N$-th Landau level. 
After the analytic continuation $i\omega_{\ell}\rightarrow\omega+i0^+$ and taking the DC limit $\omega\rightarrow0$, one obtains
\begin{equation} \label{eqn:dc_conductivity}
    \sigma_{xx}(\mu,B) = \frac{e^2}{2\hbar l_B^2} \int dE
    \left(-\frac{\partial f(E-\mu)}{\partial E}\right)
    \sum_{N,N'} |{v}^{x}_{NN'}|^2
    A_N(E,B)A_{N'}(E,B).
\end{equation}
where 
\begin{equation} \label{eqn:spectral_function}
    A_N(E,B) = -\frac{1}{\pi} \operatorname{Im}G_N(E,B).
\end{equation}
In deriving Eq.~\eqref{eqn:dc_conductivity}, we used the Landau-level degeneracy per unit area,
$\frac{1}{\mathcal{A}}\sum_X =  \frac{eB}{2\pi\hbar} = \frac{1}{2\pi l_B^2}$. 
Assuming a nearly isotropic conductivity tensor, the longitudinal resistivity is
\begin{equation} \label{eq:rho_from_sigma}
    \rho_{xx} = \frac{\sigma_{xx}}
    {\sigma_{xx}^2+\sigma_{xy}^2}.
\end{equation}
We estimate the Hall conductivity from the classical filling relation,
\begin{equation}
    \sigma_{xy} \simeq \nu\frac{e^2}{h} = \frac{ne}{B},
\end{equation}
where $n$ is the carrier density. 
In the experimentally relevant regime $|\sigma_{xy}|\gg\sigma_{xx}$, so we have $
\rho_{xx} \simeq \frac{\sigma_{xx}}{\sigma_{xy}^2} = \frac{B^2}{n^2e^2}\sigma_{xx}$.

Note that the diagonal velocity matrix elements vanish for quantized Landau levels, $v^x_{NN}=0$. 
The longitudinal conductivity is therefore dominated by transitions between neighboring Landau levels. 
Meanwhile, the spectral function $A_N(E,B)$ is peaked near the energy of the $N$th Landau level. 
Consequently, the conductivity is determined by the overlap of two neighboring spectral functions, and would be enhanced when neighboring Landau levels approach one another in energy.

A quantitative evaluation of the resistivity therefore depends on the form of the spectral function $A_N(E,B)$. 
In the following, we employ two disorder-broadening schemes to determine the spectral function. 
The first is the self-consistent Born approximation (SCBA) for short-range disorder, while the second is a simple Lorentzian broadening. 
Both schemes yield qualitatively consistent results.

\subsection{Self-consistent Born Approximation}

We estimate the disorder-averaged Green's function within the self-consistent Born approximation (SCBA). 
For short-range white-noise disorder with impurity density $n_{\mathrm{imp}}$ and strength $u_0$, and neglecting Landau level-dependent form factors, the rainbow diagrams give
\begin{equation} \label{eqn:SCBA_equations_supp}
    \Sigma_N(E,B) = \frac{n_{\mathrm{imp}}u_0^2}{2\pi l_B^2}\sum_{N'}G_{N'}(E,B), \quad 
    G_N(E,B) = \frac{1}{E-E_N-\Sigma_N(E,B)}.  
\end{equation}
For simplicity, here we assume short-range disorder with a momentum-independent correlator that is diagonal in valley. 
The sublattice and layer dependence is also omitted as the density is mainly polarized in the $A_1$ sublattice in a large displacement field. 
Within this structureless-disorder approximation, the self-energy $\Sigma_N(E)$ is independent of the Landau-level index. 
Here we denote 
\begin{equation}
    \frac{\Gamma(B)^2}{4} = \frac{n_{\mathrm{imp}}u_0^2}{2\pi l_B^2}
\end{equation}
as the scattering strength. 
Here $1/(2\pi l_B^2)$ comes from the degeneracy of each Landau level, 
as a result, $\Gamma(B)^2$ increase linearly with the magnetic field strength $|B|$. 
To characterize the strength of disorder, we introduce a temperature scale $T_D$ to denote the strength of disorder, with $T_D$ taken as the disorder strength at $B_0=1T$, i.e.
\begin{equation}
    \Gamma(B) = \pi k_B T_D \sqrt{B/B_0},\quad  B_0 = 1T
\end{equation}

When a single Landau level is well separated from all the others, the sum over $N'$ in Eq.~\eqref{eqn:SCBA_equations_supp} can be approximated by the single term $G_N$. The resulting self-consistent self-energy is
\begin{equation}
    \Sigma_N(E,B) = \frac{E-E_N}{2} - i\frac{ \sqrt{\Gamma^2-(E-E_N)^2}}{2}~,
\end{equation}
where the imaginary part of the self-energy has a semicircular profile.
In the present problem, however, crossings between Landau levels are essential, and the overlap between neighboring Landau levels cannot be neglected. 
We therefore solve Eq.~\eqref{eqn:SCBA_equations_supp} numerically while retaining all Landau levels.  
Representative convergent self-energies are shown in Fig.~\ref{fig:SCBA}. 
Note that, according to Eq.~\eqref{eqn:SCBA_equations_supp}, the self-energy is independent of the Landau-level index $N$. 
For well-separated Landau levels, the spectral function remains close to the isolated-level semicircular form. 
When two levels approach one another, their spectral functions overlap, and the magnitude of the imaginary self-energy is enhanced.

\subsection{Lorentzian broadening. }

\begin{figure}
    \centering
    \includegraphics[width=0.75\linewidth]{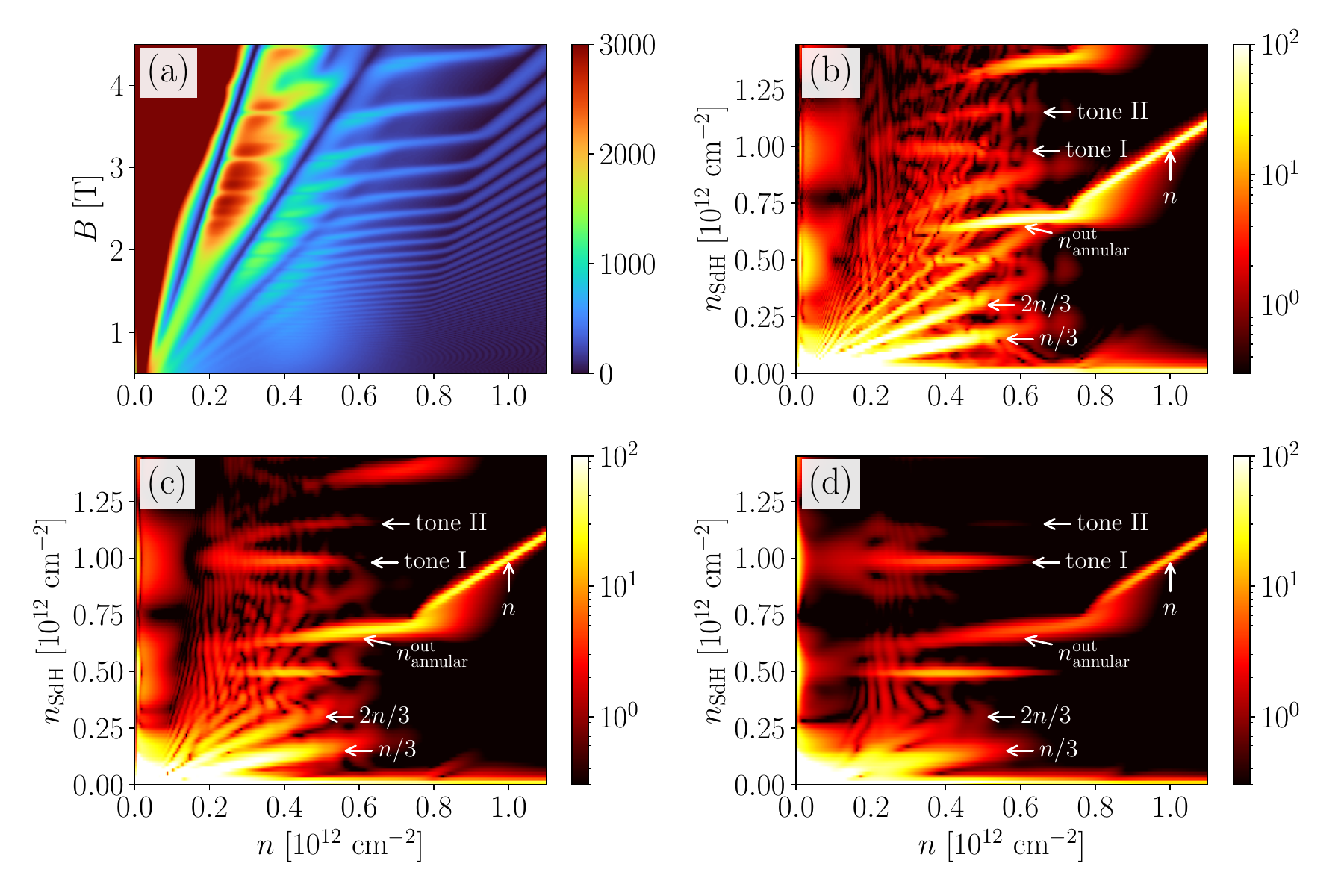}
    \caption{Longitudinal resistivity and the corresponding SdH oscillations obtained using the simple Lorentzian broadening scheme in Eq.~\eqref{eqn:Dingle}, 
    with Dingle temperature $T_D=0.05,\mathrm{K}$. All calculations are performed at $u_D=43,\mathrm{meV}$. 
    (a) Longitudinal resistivity $\rho_{xx}(n,B)$ at $T=0.5\,\mathrm{K}$. 
    (b)--(d) Fourier spectra of $\rho_{xx}(n,B)$ at $T=0.2$, $0.5$, and $1.0\,\mathrm{K}$, respectively. }
    \label{fig:Dingle}
\end{figure}

We can also consider an even simpler treatment of disorder by replacing the delta-function spectral function directly with a Lorentzian-broadened form:
\begin{equation}\label{eqn:Dingle}
    A_N(E,B) = \frac{\Gamma/\pi}{(E-E_N)^2+\Gamma^2}~,
\end{equation}
where $\Gamma=\pi k_B T_D$ and $T_D$ is the Dingle temperature. 
This corresponds to the standard phenomenological treatment of Dingle broadening \cite{dingle1952}.

In Fig.~\ref{fig:Dingle}, we show the longitudinal resistivity obtained within this approximation for $u_D=43\,\mathrm{meV}$ and $T_D=0.05\,\mathrm{K}$ at several temperatures. 
The results are qualitatively consistent with those in Fig.~\ref{fig:Resxx}, where the SCBA is employed. 
In particular, both ``multitone'' peaks remain clearly visible, although tone II exhibits a slightly different density dependence. 
We therefore conclude that the ``multitone'' structure is robust against the choice of disorder-broadening scheme.

\section{Additional Results for Different Parameters and Temperatures}\label{app:additional_results}

\begin{figure}[t]
    \centering
    \includegraphics[width=0.9\linewidth]{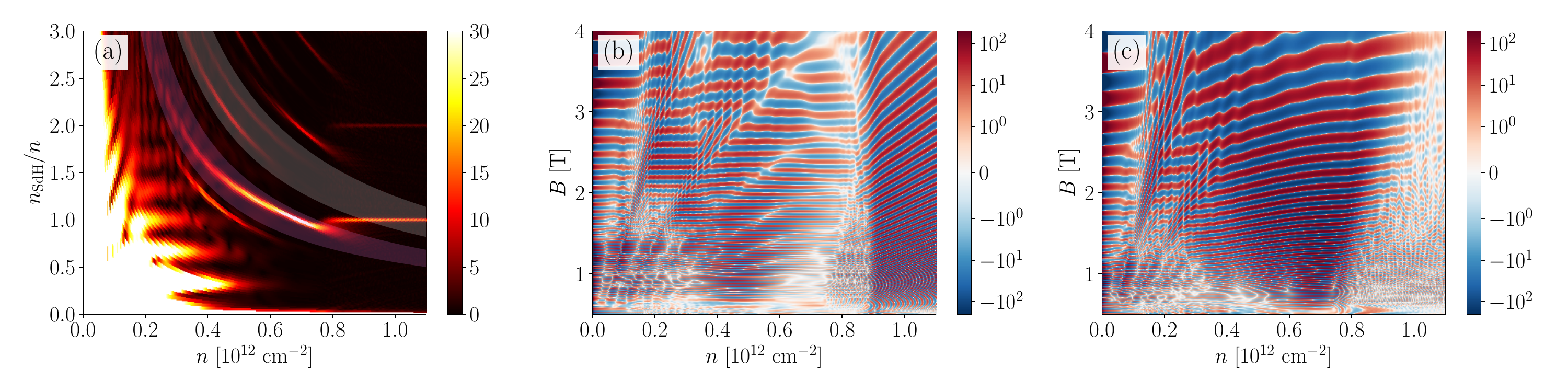}
    \caption{(a) SdH spectra ploted in the normalized SdH density $n_{\mathrm{SdH}}/n$ and linear intensity scale at $u_D=43\,\mathrm{meV}$, $T=0.5\,\mathrm{K}$, and $T_D=0.05\,\mathrm{K}$. 
    (b) and (c) show inverse Fourier reconstruction of the resistivity oscillations $\Delta\rho_{xx}$ at
    (b) Reconstruction using the ``multitone'' frequency window $0.9\times10^{12}\,\mathrm{cm}^{-2}<n_{\mathrm{SdH}}<1.25\times10^{12}\,\mathrm{cm}^{-2}$, as indicated by the shaded white regime in (a). 
    The peaks are in correspondence with the ring structures in Fig.~\ref{fig:Resxx} (a). 
    The beating pattern induced by different triplets is also seen. 
    (c) Reconstruction using the outer annular Fermi surface frequency window $0.55\times10^{12}\,\mathrm{cm}^{-2}<n_{\mathrm{SdH}}<0.75\times10^{12}\,\mathrm{cm}^{-2}$ as indicated by the shaded purple regime in (a).
    }
    \label{fig:InvSdH}
\end{figure}

\subsection{Inverse Fourier transformation. }

To gain further insight into the different oscillation components, in Fig.~\ref{fig:InvSdH} we reconstruct the longitudinal resistivity oscillations by performing an inverse Fourier transform over selected frequency windows corresponding to the ``multitone'' regime and the outer annular Fermi surface regime.
Comparison with the full resistivity map in Fig.~\ref{fig:Resxx}(a-b) reveals that:
(I) The ``multitone'' components reproduce structures consistent with the ring-like features in the resistivity map.
(II) The superposition of the two tones gives rise to the beating pattern, manifested as the modulation between neighboring Landau-level triplets.
(III) The annular Fermi surface component instead accounts for the smoother background structure.

\subsection{Temperature dependence.}

\begin{figure}[t]
    \centering
    \includegraphics[width=0.8\linewidth]{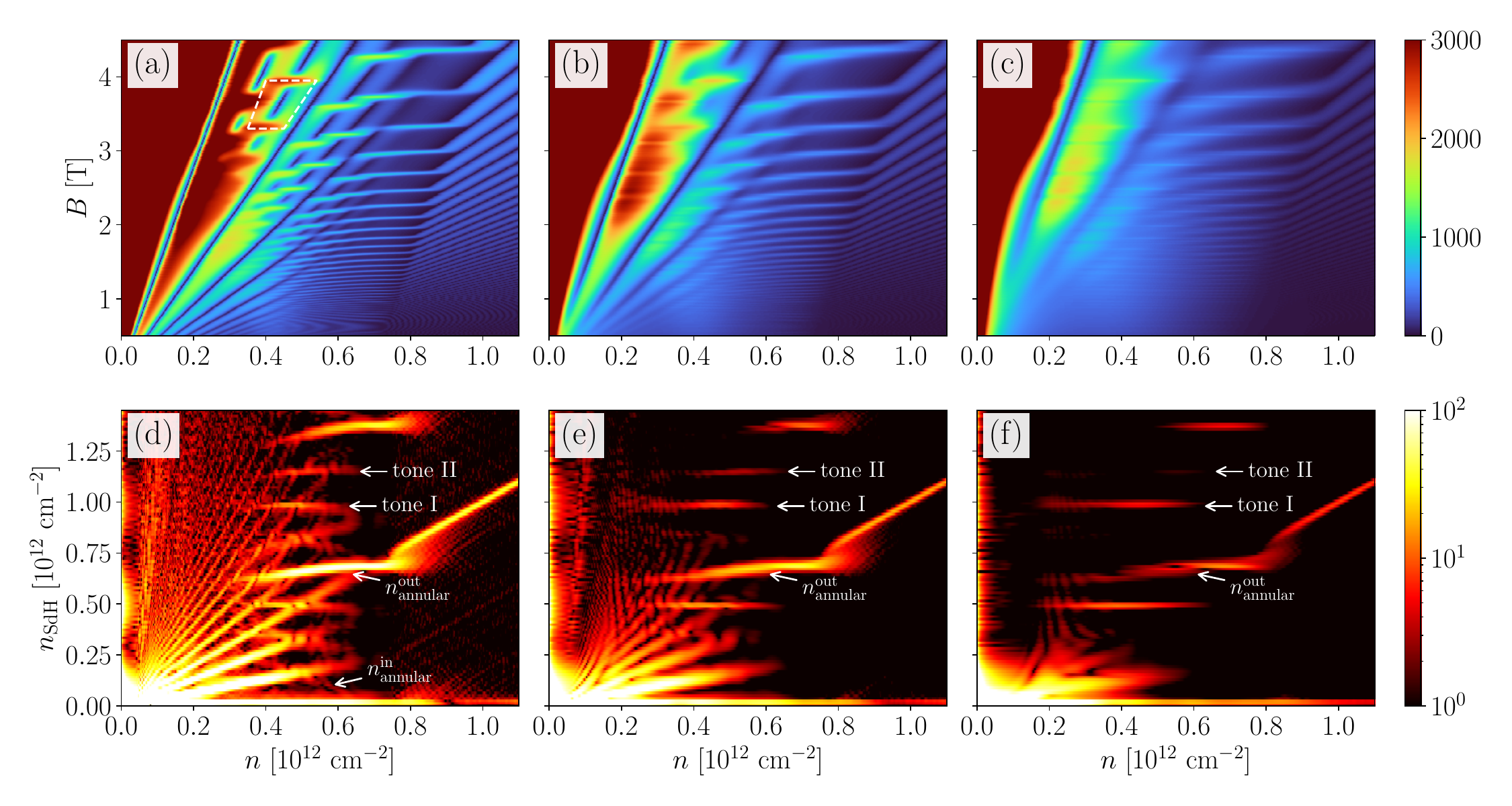}
    \caption{Temperature dependence of the longitudinal resistivity map $\rho_{xx}$ and its Fourier analysis for $u_D=43~\mathrm{meV}$ and $T_D=0.05~\mathrm{K}$. 
    (a-c) Longitudinal resistivity map $\rho_{xx}(n,B)$ for $T=0.2$K, $T=0.5$K and $T=1.0$K, respectively. 
    An example ring structure is marked in (a) by the dashed lines. 
    (d-e) The corresponding Fourier spectra of $\rho_{xx}(n,B)$ for $T=0.2$K, $T=0.5$K and $T=1.0$K, respectively. 
    As temperature increases, all the peaks become weaker, but the annular FS peaks decay much faster. 
    }
    \label{fig:rxxmap_T}
\end{figure}
\begin{figure}[t]
    \centering
    \includegraphics[width=0.8\linewidth]{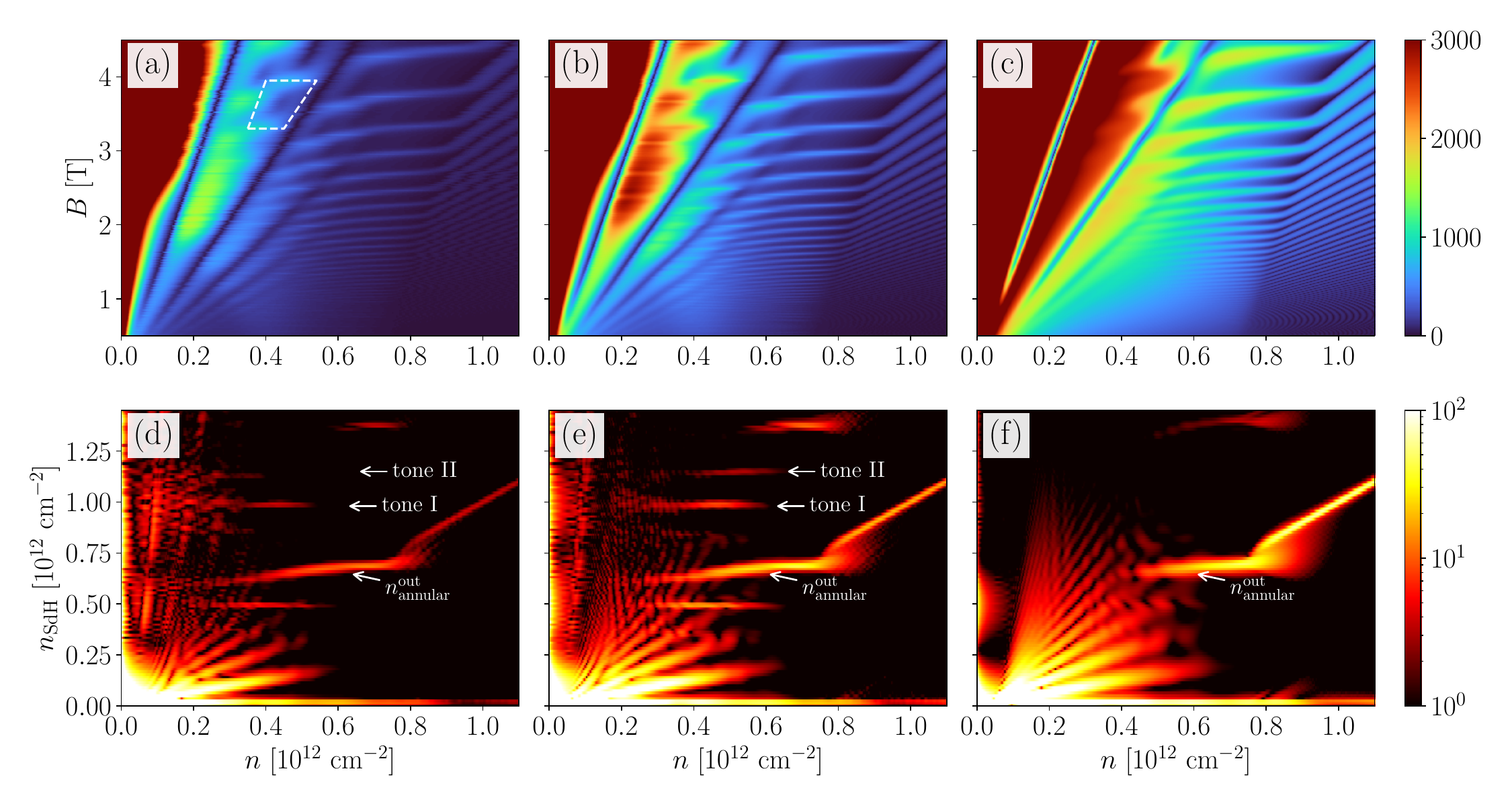}
    \caption{Disorder dependence of the longitudinal resistivity map $\rho_{xx}$ and its corresponding Fourier analysis for $u_D=43~\mathrm{meV}$ and $T=0.5~\mathrm{K}$. 
    (a-c) Longitudinal resistivity map $\rho_{xx}(n,B)$ for $T_D=0.02$K, $T_D=0.05$K and $T_D=0.20$K, respectively. 
    An example ring structure is marked in (a) by the dashed lines.  
    (d-e) The corresponding Fourier spectra of $\rho_{xx}(n,B)$ for $T_D=0.02$K, $T_D=0.05$K and $T_D=0.20$K, respectively. 
    }
    \label{fig:rxxmap_TD}
\end{figure}

In Fig.~\ref{fig:rxxmap_T}, we show the temperature evolution of the longitudinal resistivity and the corresponding SdH oscillation map at fixed $u_D=43\,\mathrm{meV}$ and $T_D=0.05\,\mathrm{K}$.
As the temperature increases, the slanted edges of the ring-like structures in the Landau-fan map are gradually washed out, whereas the two nearly horizontal segments broaden in carrier density and become more pronounced. 
Correspondingly, in the SdH oscillation spectra, the peaks associated with the annular Fermi surface fade with increasing temperature, while the ``multitone'' frequencies remain more robust towards temperature.

\subsection{Disorder strength dependence. }

In Fig.~\ref{fig:rxxmap_TD}, we show the evolution of the longitudinal resistivity and the corresponding SdH oscillation map with disorder strength at fixed $u_D=43\,\mathrm{meV}$ and $T=0.5\,\mathrm{K}$.
As the disorder strength increases, the spectral functions $A_N(E,B)$ broaden, increasing the overlap between neighboring Landau levels and initially enhancing the SdH oscillation peaks. 
With further increasing disorder, however, the Landau-level structure is washed out, and the oscillation amplitudes decrease. 
As a result, both the peaks associated with the annular Fermi surface and the ``multitone'' frequencies are suppressed at stronger disorder, with the ``multitone'' components being more strongly affected. 
The annular Fermi surface component remains more prominent throughout the disorder range considered.

\subsection{Displacement field dependence. }

\begin{figure}[t]
    \centering
    \includegraphics[width=0.99\linewidth]{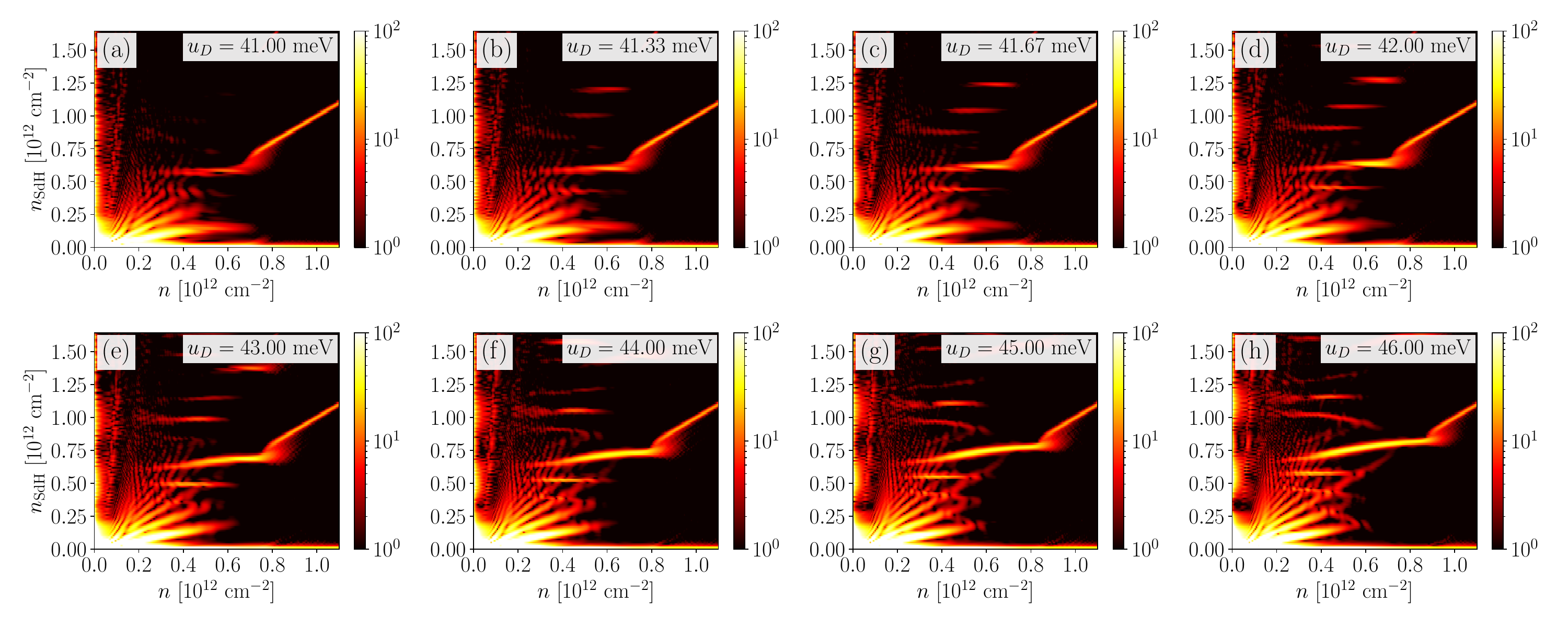}
    \caption{
    SdH oscillations of longitudinal resistivity for displacement fields 
    (a) $u_D = 41.00~\mathrm{meV}$,
    (b) $u_D = 41.33~\mathrm{meV}$,
    (c) $u_D = 41.67~\mathrm{meV}$,
    (d) $u_D = 42.00~\mathrm{meV}$,
    (e) $u_D = 43.00~\mathrm{meV}$,
    (f) $u_D = 44.00~\mathrm{meV}$,
    (g) $u_D = 45.00~\mathrm{meV}$,
    and (h) $u_D = 46.00~\mathrm{meV}$, respectively. 
    All calculations are performed at $T=0.5~\mathrm{K}$ with a disorder broadening corresponding to $T_D=0.05~\mathrm{K}$.
    }
    \label{fig:ud_supp}
\end{figure}

Finally, in Fig.~\ref{fig:ud_supp}, we present additional results for displacement fields ranging from $u_D=41.00~\mathrm{meV}$ to $u_D=46.00~\mathrm{meV}$. 
The ``multitone'' structure remains robust over a broad range of displacement fields, provided that VHSs separates the three-pocket Fermi surface from the annular Fermi surface.
For all displacement fields at which the ``multitone'' structure is observable, it persists over a broad range of carrier densities around the VHS. 
The density range over which at least one of the ``multitone'' frequencies can be resolved is marked in the DOS map in Fig.~\ref{fig:FS}(d). 
This regime closely follows the top-left VHS separating the three-pocket and annular Fermi surfaces. 
Remarkably, the ``multitone'' structure extends over a broader density range than might be expected from the zero-field Fermi-surface geometry. 
For $u_D\leq42~\mathrm{meV}$, it remains visible even in a regime where the zero-field Fermi surface is singly connected.

The ``multitone'' frequencies increase slightly with increasing $u_D$, consistent with the experimental observations.
For $u_D\geq43~\mathrm{meV}$, however, an additional frequency appears below tone I. 
This frequency corresponds to the combined area $A_{\mathrm{in}}+A_{\mathrm{out}}$ of the annular Fermi surface. 
As $u_D$ increases, spectral weight is gradually transferred from tone II to this additional frequency.

\subsection{Magnetic field dependence. }

\begin{figure}[t]
    \centering
    \includegraphics[width=0.99\linewidth]{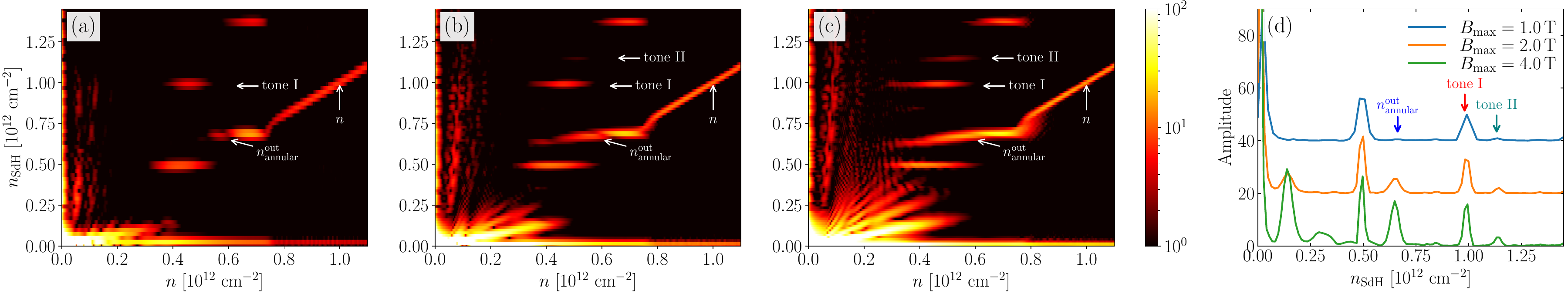}
    \caption{
    Dependence of the SdH Fourier spectra on the magnetic-field window.
    All the datas are Fourier transformation of the same data of Fig.~\ref{fig:Resxx}(a) with $u_D=43~\mathrm{meV}$, $T=0.5~\mathrm{K}$ and $T_D=0.05~\mathrm{K}$. 
    All spectra are obtained by Fourier transforming the same resistivity data shown in Fig.~\ref{fig:Resxx}(a), calculated at $u_D=43~\mathrm{meV}$, $T=0.5~\mathrm{K}$, and $T_D=0.05~\mathrm{K}$, 
    using different magnetic-field ranges: (a) $B=0.5$--$1.0~\mathrm{T}$, (b) $B=0.5$--$2.0~\mathrm{T}$, and (c) $B=0.5$--$4.0~\mathrm{T}$. 
    The magnetic-breakdown-induced features become more pronounced as the upper bound of the magnetic-field window increases.
    (d) The fixed density line cuts at $n=0.45\times 10^{12}~\mathrm{cm}^{-2}$ for (a)--(c), respectively. 
    }
    \label{fig:SdH_Bs}
\end{figure}

Since magnetic breakdown becomes stronger with increasing magnetic field, we examine in Fig.~\ref{fig:SdH_Bs} the dependence of the SdH spectra on the magnetic-field window. 
As the upper bound $B_{\max}$ increases, both ``multitone'' features become more pronounced and extend over a broader range of carrier density. 
In particular, tone II is nearly absent for $B_{\max}=1~\mathrm{T}$, whereas tone I remains clearly visible and is considerably more robust at low fields. 
In the main text, we use a broader magnetic-field window, $B=0.5$--$8.0~\mathrm{T}$ for a clearer visibility. 
Nevertheless, the essential features are already well developed within the experimentally relevant range $B=0.5$--$4.0~\mathrm{T}$. 

\section{Semiclassical quantization of an $M$-pocket toy model}

\begin{figure}[t]
    \centering
    \includegraphics[width=0.75\linewidth]{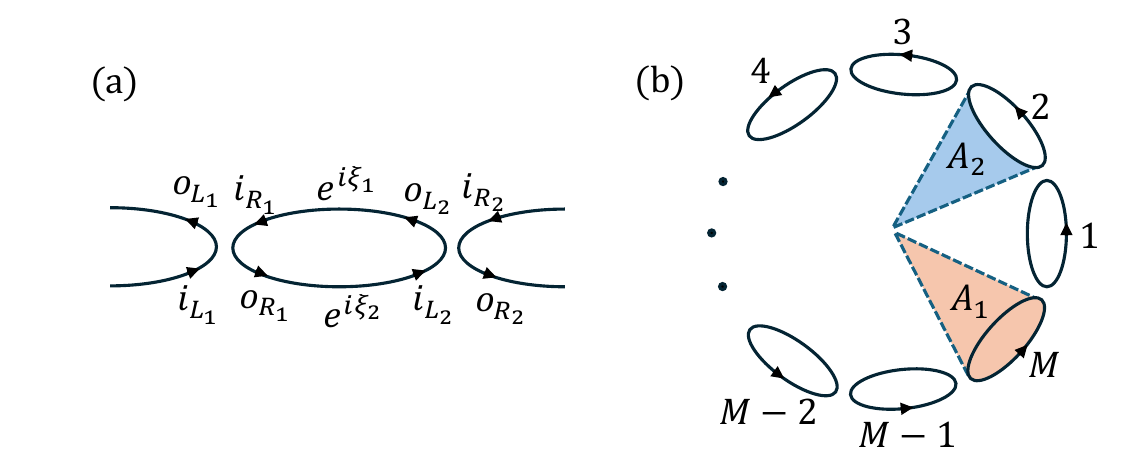}
    \caption{Schematic illustration of the finite magnetic-breakdown network. 
    (a) Two adjacent junctions connected by a Fermi pocket. 
    We use $i_{Rk}$ and $i_{Lk}$ to denote the incoming semiclassical wave packets, and $o_{Rk}$ and $o_{Lk}$ to denote the outgoing wave packets. 
    The factors $e^{i\xi_1}$ and $e^{i\xi_2}$ represent the phases accumulated along the upper and lower halves of the pocket, respectively.
    (b) A ring consisting of $M$ Fermi pockets separated by VHSs.
    The areas enclosed by the outer and inner half-pocket trajectories with the origin are denoted by $A_1$ and $A_2$, respectively.
    The separation between neighboring pockets is assumed to be small.
    }
    \label{fig:MB_network}
\end{figure}

In this section, we show a semiclassical derivation of the Magnetic breakdown effect for an $M$-pocket model as illustrated in Fig.~\ref{fig:MB_network}. 
The braiding Landau level structure is analyzed starting from both the separated pockets limit and the annular Fermi surface limit. 

Magnetic breakdown occurs when the cyclotron motion of an electron in momentum space 
becomes nonadiabatic within a narrow region where two segments of Fermi surface are close to each other \cite{pippard1962a,pippard1964,chambersMB1966}. 
Motion away from each junction can still be treated semiclassically. 
Each junction is described by a scattering matrix, whereas propagation along a semiclassical orbit contributes a dynamical phase.
Repetition of this elementary structure in a periodic system leads to an effective orbit-network model. 
For an infinite network, the otherwise dispersionless Landau levels can acquire a finite bandwidth and form magnetic minibands. 

In the present problem, we instead consider the Fermi surface forms a finite ring of $M$ symmetry-related pockets. 
Semiclassically, this system can be viewed as a sequence of coupled Fabry--P\'erot interferometers, as illustrated in Fig.~\ref{fig:MB_network}.

We start from a single connection point between two neighboring orbit segments. 
For the $k$-th junction, the incoming and outgoing amplitudes are related by the scattering matrix
\begin{equation} \label{eqn:single_scattering}
    \begin{pmatrix}
        \psi_{oL_k}\\\psi_{oR_k} 
    \end{pmatrix} = \mathcal{T}
    \begin{pmatrix}
        \psi_{iL_k}\\\psi_{iR_k} 
    \end{pmatrix},
    \qquad
    \mathcal{T} = \begin{pmatrix}
        t_{11} & t_{12}\\
        t_{21} & t_{22}
    \end{pmatrix}.
\end{equation}
where $i,o$ denote incoming and outgoing amplitudes, while $L,R$ label the two sides of the junction.
$\mathcal{T}$ is the unitary scattering matrix. 
It is convenient to rewrite Eq.~\eqref{eqn:single_scattering} in transfer-matrix form,
\begin{equation}\label{eqn:single_transfer}
    \begin{pmatrix}
        \psi_{oR_k}\\\psi_{iR_k} 
    \end{pmatrix} = \mathcal{M}
    \begin{pmatrix}
        \psi_{oL_k}\\\psi_{iL_k} 
    \end{pmatrix}
    ,\qquad
    \mathcal{M} = \begin{pmatrix}
        t_{22}/t_{12} & t_{21}-t_{11}t_{22}/t_{12} \\
        1/t_{12} & -t_{11}/t_{12} 
    \end{pmatrix}~,
\end{equation}
where $\mathcal{M}$ is the transfer matrix.  

We next include the phases accumulated along the two semiclassical arcs forming one lens. 
Denoting these phases by $\xi_1$ and $\xi_2$, the transfer matrix from the $k$th junction to the $(k+1)$th junction is described by
\begin{equation}
    \begin{pmatrix}
        \psi_{oL_{k+1}}\\\psi_{iL_{k+1}} 
    \end{pmatrix} = 
    \mathcal{P}
    \begin{pmatrix}
        \psi_{oR_{k}}\\\psi_{iR_k} 
    \end{pmatrix}
    ,\qquad 
    \mathcal{P} =
    \begin{pmatrix}
        0&e^{-i\xi_1}\\ e^{i\xi_2}&0
    \end{pmatrix}.
\end{equation}
The gauge-dependent phases $\xi_1$ and $\xi_2$ contain both the area-dependent orbital phases and constant phase offsets associated with the Maslov phase. 
For the $M$-pocket model in Fig.~\ref{fig:MB_network}(b), we take the gauge 
\begin{equation}
    \xi_1 =l_B^2 A_1+\pi/M + \pi/2
    ,\qquad
    \xi_2 = -l_B^2A_2 - \pi/M + \pi /2~,
\end{equation}
where $A_1$ and $A_2$ are the areas enclosed by the two arcs and a given orientation point as indicated in Fig.~\ref{fig:MB_network} (b). 
The additional $\pm \pi/M+\pi/2$ phases are included to obtain the Maslov phase for any closed loop and are not essential to our subsequent derivation. 
Additional phases contributed by the Berry connection and orbital moment are omitted here for simplicity. 

The transfer matrix across one full elementary cell is therefore
\begin{equation} \label{eqn:cell_transfer}
    \mathcal{U} = \mathcal{P}\mathcal{M} = \begin{pmatrix}
        e^{-i\xi_1}/t_{12} & -e^{-i\xi_1} t_{11}/t_{12} \\
        e^{i\xi_2} t_{22} / t_{12} & e^{i\xi_2} (t_{21} - t_{11}t_{22}/t_{12})~.
    \end{pmatrix}
\end{equation}

For a ring consisting of $M$ equivalent cells, 
the quantization condition Eq.~\eqref{eqn:OBS} should be replaced by the single-valuednes condition $\mathcal{U}^{M}\Psi_m=\Psi_m$, 
or equivalently 
\begin{equation}
\label{eqn:network_quan}
    \mathcal{U}\Psi_m=e^{\frac{2\pi i m}{M}}\Psi_m
    \quad \Longleftrightarrow\quad 
    \det\left[\mathcal{U}-e^{\frac{2\pi i m}{M}}\mathbbm{1}\right]=0,
    \qquad
    m=0,1,\ldots,M-1, 
\end{equation}
where the integer $m$ labels the discrete angular-momentum sector of the finite orbit network.

We now specialize to a case where the junction can be approximated by a quadratic saddle point, 
where the scattering matrix is \cite{alexandradinataModernMB2018}
\begin{equation} \label{eqn:saddle_scattering}
    \mathcal{T} = \begin{pmatrix}
        T&R\\R&T
    \end{pmatrix}
    ,\qquad
    T = \frac{e^{i\phi}}
    {\sqrt{1+\exp(\pi\varepsilon)}},
    \qquad
    R =
    \frac{-ie^{i\phi}}
    {\sqrt{1+\exp(-\pi\varepsilon)}}~. 
\end{equation}
Here, $\varepsilon = \sqrt{m_1m_2}(E-E_{\mathrm{VHS}})l_B^2/\hbar$ is a dimensionless parameter.
$E_{\mathrm{VHS}}$ is the energy of VHS, and $m_{1,2}$ are the two local curvature masses of the saddle point. 
The scattering phase $\phi$ is
\begin{equation}
    \phi(\varepsilon) =
    \arg\Gamma_E\left(\frac{1}{2}+i\varepsilon\right)
    -\varepsilon\ln|\varepsilon|+\varepsilon~,
\end{equation}
where $\Gamma_E(z)$ denotes the Euler gamma function.
$\phi(\varepsilon)$ approaches zero for both $\varepsilon=0$ and $|\varepsilon|\rightarrow\infty$.

Substituting Eq.~\eqref{eqn:saddle_scattering} into Eq.~\eqref{eqn:network_quan} gives the quantization condition as 
\begin{equation} \label{eqn:MB_quantization}
    |R| \cos\left( \frac{2\pi m}{M}+\frac{\xi_1-\xi_2}{2} \right) =
    \sin\left( \frac{\xi_1+\xi_2}{2}+\phi \right), \qquad
    m = 0,1,\ldots M-1.
\end{equation}
Solving Eq.~\eqref{eqn:MB_quantization} gives all the quantized Landau level energies. 
Here we consider two limiting cases where the magnetic breakdown effect can be treated perturbatively. 

\subsection{Separated $M$ pockets limit.}

First, when $\varepsilon\rightarrow-\infty$, for which $|T|\rightarrow1$, $|R|\rightarrow0$, and $\phi\rightarrow0$, 
the problem reduces to $M$ decoupled small pockets. 
The quantization condition Eq.~\eqref{eqn:MB_quantization} reduces to
\begin{equation}
    \sin\left( \frac{\xi_1+\xi_2}{2} \right)=0,
\end{equation}
and hence
\begin{equation} \label{eqn:combined_orbit}
    l_B^2(A_1-A_2)
    =(2N-1)\pi ,
    \qquad N\in\mathbb{Z}.
\end{equation}
Only the total phase accumulated around a single closed pocket enters the quantization condition. 
The sector label $m$ drops out, and all $M$ angular-momentum sectors are degenerate.

For small but finite $|R|$, the $M$ degenerate branches split. 
We write $A_1-A_2=(2N-1)\pi/l_B^2+\delta A_{N,m}$,
with $ |l_B^2 \delta A_{N,m}|\ll1 $.
Expanding Eq.~\eqref{eqn:MB_quantization} to leading order in $|R|$ gives
\begin{equation}
   l_B^2 \delta A_{N,m} = 2|R| \cos\left(\frac{2\pi m+\Phi}{M} + N\pi\right) 
    +\mathcal{O}(|R|^2), \qquad m = 0, 1,\ldots M-1.
\end{equation}
In the $M=3$ case, the above result is equivalent to the effective three-band model Eq.~\eqref{eqn:es_MB} up to a constant phase factor $N\pi$. 
The phase $\Phi$ in Eq.~\eqref{eqn:es_MB} in the semi-classical limit is therefore 
\begin{equation}\label{eqn:crossingPhi}
    \Phi=\frac{M}{2}(\xi_1-\xi_2) = \frac{M}{2}l_B^2(A_1+A_2)+\pi~ ,
\end{equation}
which is the average area of the outer and inner pockets of the annular Fermi surface.  
The phase $\Phi$ can be interpreted as the total flux-induced phase accumulated around a one-dimensional $M$-site tight-binding ring with periodic boundary conditions.

We note that Landau-level crossings discussed in the main text are not restricted to the $M=3$ case. 
For general $M\geq 3$, similar Landau-level crossings  occur when $\Phi$ is an inter multiple of $\pi$, or equivalently: 
\begin{equation}\label{eqn:crossingM}
    M l_B^2(A_1+A_2) = 0\ (\mathrm{mod}\ 2\pi)~. 
\end{equation}
These Landau-level braiding structures are therefore expected to produce anomalous oscillation frequencies associated with the area $M(A_1+A_2)$, 
accompanied by a beating-induced splitting controlled by $A_1-A_2$.

\subsection{Relating the ``multitone'' frequency to the Fermi surface geometry. }

With Eqs.~\eqref{eqn:crossingPhi} and \eqref{eqn:crossingM}, we are ready to relate the ``multitone'' frequencies to the Fermi surface geometry. 
Here we focus on the $M=3$ case relevant to the R$n$G, and denote the inner and outer area of the annular Fermi surface as $A_{\mathrm{in}}$  and $A_\mathrm{out}$. 
In the weak magnetic breakdown regime with three separated pockets, their area are determined by the quantization condition: 
\begin{equation}
    l_B^2(A_{\mathrm{out}} - A_{\mathrm{in}})/3 = 2\pi N-\pi~. 
\end{equation}
Here we omit the geometry phases contributed by the Berry connection, which is not essential in the current analysis. 
To leading order in the dimensionless parameter $x=nl_B^2$, we expand the areas as 
\begin{equation}
    A_{\mathrm{out}} = A_0 + A_{\mathrm{out}}' x+O(x^2),\qquad 
    A_{\mathrm{in}} = A_0 + A_{\mathrm{in}}'x+O(x^2)~,
\end{equation}
where $A'_{\mathrm{out}}\equiv \mathrm{d}A_{\mathrm{out}}/\mathrm{d}x$ and $A'_{\mathrm{in}}\equiv \mathrm{d}A_{\mathrm{in}}/\mathrm{d}x$. 
Note that $A_{\mathrm{out}}-A_{\mathrm{in}}$ is proportional to the total carrier density and vanishes at $x=0$. 

With these relations, the condition of Landau level crossing $l_B^2(A_{\mathrm{out}}+A_{\mathrm{in}})=0\,(\mathrm{mod}\ 2\pi)$ becomes 
\begin{equation}
    l_B^2(A_{\mathrm{out}}+A_{\mathrm{in}}) = 2l_B^2A_0+ (2\pi N-\pi) \frac{3(A_{\mathrm{out}}'+A_{\mathrm{in}}')}{A_{\mathrm{out}}'-A_{\mathrm{in}}'} + O(x^2)~. 
\end{equation}
The second term is independent of the magnetic field $B$. 
The field-dependent part of the phase is therefore controlled by $2A_0$, while the remaining term contributes only a constant phase offset. 
Consequently, the area inferred from the Fourier-transformed oscillations is $2A_0$, rather than the instantaneous value $A_{\mathrm{out}}+A_{\mathrm{in}}$. 
The value of $A_0$ can be extracted from the zero-field intercepts of linear fits to $A_{\mathrm{out}}$ and $A_{\mathrm{in}}$ as functions of $nl_B^{2}$, as indicated as an example in Figs.~\ref{fig:Resxx} (c) and \ref{fig:Resxx} (d).

\subsection{Annular Fermi sea limit. }

We next consider the opposite limit with $\varepsilon\rightarrow +\infty$, where $|T|\rightarrow0$, $|R|\rightarrow1$ and $\phi\rightarrow 0$. 
Here, the $M$ pockets are connected to each other to form an annular Fermi sea with outer and inner Fermi pockets. 
Eq~\eqref{eqn:MB_quantization} becomes
\begin{equation}
    \cos\left( \frac{2\pi m}{M}+\frac{\xi_1-\xi_2}{2} \right) =
    \sin\left( \frac{\xi_1+\xi_2}{2} \right).
\end{equation}
This solution separates into two independent branches,
\begin{align}
    Ml_B^2A_1 &= -\pi -2\pi m + 2\pi M N, \\
    Ml_B^2A_2 &= -\pi + 2\pi m+ 2\pi M N'.
\end{align}
The two equations describe the independent semiclassical quantization of the inner and outer cyclotron orbits of the annular Fermi surface, enclosing the areas $MA_1$ and $MA_2$, respectively.

To understand how magnetic breakdown modifies these solutions, consider a small deviation from the uncoupled point,
$A_i=A_{i,0}+\delta A_i$.  
Using $|R|=\sqrt{1-|T|^2}=1-|T|^2/2+\mathcal{O}(|T|^4)$, the leading corrections to the two branches are
\begin{align}
    l_B^2\delta A_1 &=
    -\frac{|T|^2}{2} \tan\left( \frac{l_B^2(A_1-A_2)+\pi}{2} \right) +\mathcal{O}(|T|^4), \\
    l_B^2 \delta A_2 &= -\frac{|T|^2}{2}
    \tan\left( \frac{l_B^2(A_1-A_2)+\pi}{2} \right) +\mathcal{O}(|T|^4).
\end{align}
These corrections have two implications. 
First, the $\tan$ correction indicates that the perturbative correction becomes singular whenever $l_B^2(A_1-A_2) = 0\ (\mathrm{mod}\ 2\pi)$.
Near these points, the correction is no longer small, indicating that a branch must reconnect with a neighboring quantization branch rather than continue perturbatively.
Second, the strength of Magnetic breakdown is periodic on the small pocket area $l_B^2(A_{1}-A_2)$. 
Note that the condition for the outer and inner Landau levels to be degenerate with each other is $Ml_B^2(A_1-A_2) = 0 \ (\mathrm{mod}\ 2\pi)$. 
The singularity condition therefore selects one out of every $M$ crossings between the uncoupled inner- and outer-orbit Landau levels to become an avoided crossing once magnetic breakdown is included.

\end{document}